\documentclass[fleqn,3p,onecolumn,11pt,nofootinbib,showkeys,showpacs]{revtex4-1}

\usepackage[small,justification=centerlast]{caption}
\usepackage{graphicx}
\usepackage{bm}
\usepackage{mathrsfs}
\usepackage[latin1]{inputenc}
\usepackage{array}
\usepackage{psfrag}
\usepackage{dsfont}
\usepackage{epsfig}
\usepackage{titletoc}
\usepackage{float}   
\usepackage{wrapfig} 
\usepackage{amsmath}\usepackage{amssymb,stmaryrd}
\usepackage{array}
\usepackage{epsfig}
\usepackage{cancel}
\usepackage[dvips]{feynmp}
\usepackage{upgreek}
\usepackage{subfig}
\usepackage{tabularx}
\usepackage{xcolor}
\usepackage{units}
\usepackage{textcomp}
\usepackage{wasysym}
\usepackage{dutchcal}

\DeclareMathAlphabet{\standardcal}{OMS}{zplm}{m}{n}
\newcommand{\Qs}{\standardcal{Q}}
\def\ring#1{{\mathaccent'27 #1}}

\usepackage{textfit} 

\usepackage{hyperref}

\renewcommand{\eqref}[1]{\mbox{Eq.~(\ref{#1})}}
\newcommand{\tabref}[1]{\mbox{Tab.~\ref{#1}}}

\newcommand{\secref}[1]{\mbox{Sec.~\ref{#1}}}

\newcommand{\appref}[1]{\mbox{App.~\ref{#1}}}

\begin{document}

\title{Classical Lagrangians and Finsler structures \\ for the nonminimal fermion sector of the Standard-Model Extension}

\author{M. Schreck} \email{marco.schreck@ufma.br}
\affiliation{Departamento de F\'{i}sica, Universidade Federal do Maranh\~{a}o \\
65080-805, S\~{a}o Lu\'{i}s, Maranh\~{a}o, Brazil
}

\begin{abstract}
This article is devoted to finding classical point-particle equivalents for the fermion sector of the nonminimal Standard-Model Extension (SME).
For a series of nonminimal operators, such Lagrangians are derived at first order in Lorentz violation using the algebraic concept of Gr\"{o}bner
bases. Subsequently, the Lagrangians serve as a basis for reanalyzing the results of certain kinematic tests of Special Relativity that were
carried out in the last century. Thereby, a number of new constraints on coefficients of the nonminimal SME is obtained. In the last part of
the paper we point out connections to Finsler geometry.
\end{abstract}

\maketitle

\newpage
\setcounter{equation}{0}
\setcounter{section}{0}
\renewcommand{\theequation}{\arabic{section}.\arabic{equation}}

\section{Introduction}

The quest for a violation of {\em CPT}- and Lorentz invariance in nature keeps going on. Violations of these fundamental symmetries have evaded
their experimental detection so far, which shows that they are good symmetries in the range of atomic energies to ultra-high energies of
cosmic rays. However, there is no compelling physical reason why these symmetries should be valid to the highest energies
imaginable, i.e., the Planck scale. Indeed, such violations are motivated by various prototypes of fundamental theories such as string theory
\cite{Kostelecky:1988zi,Kostelecky:1991ak,Kostelecky:1994rn}, loop quantum gravity \cite{Gambini:1998it,Bojowald:2004bb}, noncommutative
spacetimes \cite{AmelinoCamelia:1999pm,Carroll:2001ws}, spacetime foams \cite{Klinkhamer:2003ec,Bernadotte:2006ya,Hossenfelder:2014hha},
quantum field theories on spacetimes with nontrivial topology \cite{Klinkhamer:1998fa,Klinkhamer:1999zh}, and Ho\v{r}ava-Lifshitz gravity
\cite{Horava:2009uw}.

The Standard-Model Extension (SME) is a model-independent framework providing a parameterization of all possible {\em CPT}- and Lorentz-violating
operators that may be present \cite{Colladay:1998fq}. This does not only concern field operators of mass dimension 3 or 4, which are comprised
in the minimal SME, but also those of higher dimension being part of the nonminimal SME \cite{Kostelecky:2009zp,Kostelecky:2011gq,Kostelecky:2013rta}.
Its field-theory description allows us to compute how measurable quantities are affected by Lorentz violation, which is mandatory for precise tests.
Based on a theorem by Greenberg \cite{Greenberg:2002uu}, all possible {\em CPT}-violating operators are included in the SME. Theoretical questions
related to the SME at tree-level in the Standard-Model couplings were tackled in \cite{Kostelecky:2000mm,oai:arXiv.org:hep-ph/0101087,Casana:2009xs,
Casana:2010nd,Klinkhamer:2010zs,Schreck:2011ai,Casana:2011fe,Hohensee:2012dt,Schreck:2013gma,Schreck:2013kja,Schreck:2014qka,Colladay:2014dua,
Casana:2014cqa,Albayrak:2015ewa} where \cite{Jackiw:1999yp,Chung:1999pt,PerezVictoria:1999uh,PerezVictoria:2001ej,Kostelecky:2001jc,Altschul:2003ce,
Altschul:2004gs,Colladay:2006rk,Colladay:2007aj,Colladay:2009rb,Gomes:2009ch,Ferrero:2011yu,Scarpelli:2013eya,Santos:2014lfa,Santos:2015koa}
deal with issues connected to quantum corrections. To date, various
minimal and a couple of nonminimal coefficients have been searched for in experiments. A yearly updated compilation of recent bounds can be
found in \cite{Kostelecky:2008ts} showing that both the field is cutting-edge and the community is very active.

The SME does not only parameterize {\em CPT}- and Lorentz violation for particle fields but it includes Lorentz violation in gravity as well
\cite{Kostelecky:2003fs}. Phenomenological studies were performed in \cite{Bailey:2006fd,Kostelecky:2007kx,Kostelecky:2008in,Bailey:2009me,
Kostelecky:2010ze,Tasson:2010nr,Tasson:2012nx,Tasson:2012au,Bailey:2014bta,Long:2014swa,Kostelecky:2015dpa} and a number of theoretical questions
were tackled in
\cite{Bonder:2013sca,Bonder:2015maa}. One of the most important results obtained for the SME gravity sector is a no-go theorem proven in
\cite{Kostelecky:2003fs}. The latter says that explicit Lorentz violation is incompatible with the geometry of curved spacetime. That shows
up as a clash between the conversation law of the energy-momentum tensor of matter and the Bianchi identities of the curved spacetime manifold
considered. Physically, it means that momentum cannot be transferred between fields and an explicitly Lorentz-violating background since the
latter does not have any dynamics.

One possible solution is to consider spontaneous Lorentz violation \cite{Kostelecky:1989jp,Kostelecky:1989jw,Bluhm:2008yt,Hernaski:2014jsa,Bluhm:2014oua,Maluf:2014dpa},
which is dynamical, or to work within a more general geometric setting than Riemannian geometry. A promising extension might be Finsler geometry.
The latter is based on a generalized path length functional that additionally depends on the angle between the line interval and an intrinsic
preferred direction on the manifold \cite{Finsler:1918,Cartan:1933,Bao:2000,Antonelli:1993}. The connection between the SME and Finsler
geometry has been demonstrated and explored in a series of articles. The starting point is a connection made between the field-theory
description of the SME and the Lagrangian of a classical, relativistic, pointlike particle \cite{Kostelecky:2010hs}. This procedure was
applied to obtain such Lagrangians for various sectors of the minimal SME \cite{Kostelecky:2010hs,Colladay:2012rv,Schreck:2014ama,Russell:2015gwa}
and for a single coefficient of the nonminimal SME \cite{Schreck:2014hga}. It was shown that by a Wick rotation these classical Lagrangians
are connected to Finsler structures \cite{Kostelecky:2011qz,Kostelecky:2012ac}. In this context, the well-known Randers space plays a
role and two novel, interesting classes of Finsler spaces were found, which are known as b space \cite{Kostelecky:2011qz} and the bipartite
spaces \cite{Kostelecky:2012ac,Silva:2013xba}. Reference \cite{Foster:2015yta} provides some models in mechanics and electrodynamics
related to b space and certain aspects of its desingularization are discussed in \cite{Colladay:2015wra}. The article \cite{Schreck:2015dsa}
explores classical analogs of Lorentz-violating photons, which are rays that satisfy the eikonal equation. Furthermore, it demonstrates
connections to Finsler geometry that are applied to describe the propagation of Lorentz-violating rays in gravitational backgrounds.
In the recent paper \cite{Silva:2015ptj} a scalar field theory is coupled to a Randers spacetime and its physical implications are
analyzed. Last but not least, there has been a couple of works applying Finsler geometry to modify relativity, gravity, and particle physics,
amongst them \cite{Brandt:1998cw,Brandt:2000ph,Perlick:2005hz,Girelli:2006fw,Gibbons:2008zi,Skakala:2008kf,Pfeifer:2011tk,Pfeifer:2011ve,
Caponio:2011dm,Pfeifer:2012mb,Chang:2012gr, Zhang:2012hi,Lammerzahl:2012kw,Itin:2014uia}. However, these articles are not directly related
to the SME, although it may be possible to map certain of the modifications to SME coefficients.

Our knowledge about Finsler structures in connection to Lorentz violation is mainly restricted to the minimal SME. Therefore, the current
article is devoted to obtaining classical point-particle analogs within its nonminimal version. The paper is organized as follows.
Section \ref{sec:classical-lagrangians-procedure} reviews the procedure of how to connect the SME field-theory language to the
description of particles in classical physics.
Thereby we introduce the algebraic concept of Gr\"{o}bner bases, which will be a formidable tool to obtain classical Lagrangians in the
nonminimal SME. Section \ref{sec:classical-lagrangians-nonminimal-sme} demonstrates how such a Lagrangian is computed for a particular
nonminimal operator at first order in Lorentz violation. Additionally, Lagrangians for other operators are derived and presented.
In \secref{sec:experimental-sensitivity} we restrict these Lagrangians to their isotropic subset of coefficients and investigate
the modified kinematics of such a classical particle. These results at hand enable us to obtain a series of new constraints for
the nonminimal SME, based on experimental results of kinematic tests of Special Relativity. Last but not least, in
\secref{sec:finsler-structures-properties} it is demonstrated how the classical Lagrangians are connected to Finsler geometry.
We obtain a generic Finsler structure and investigate some of its mathematical properties. Finally, all results are concluded
on in \secref{sec:conclusions}. Large formulas and calculational details worth mentioning are relegated to Appx.~\ref{sec:obtaining-lagrangian-m5}
-- \ref{sec:cartan-matsumoto-torsion}. Throughout the paper, natural units with $\hbar=c=1$ will be used unless otherwise stated.
Furthermore, Greek indices always run from $0\dots 3$ whereas Latin indices run from $1\dots 3(4)$.

\section{Classical Lagrangians of the SME fermion sector}
\label{sec:classical-lagrangians-procedure}

The current article is based on the SME fermion sector that was studied in \cite{Kostelecky:2013rta} extensively. It is described by
the following action:
\begin{subequations}
\label{eq:action-fermionic-sme}
\begin{align}
S&=\int \mathrm{d}^4x\,\mathcal{L}\,, \\[2ex]
\mathcal{L}&=\frac{1}{2}\overline{\psi}(\gamma^{\mu}\mathrm{i}\partial_{\mu}-m_{\psi}+\widehat{\Qs})\psi+\text{H.c.}
\end{align}
\end{subequations}
All fields are defined in Minkowski spacetime with the metric $\eta_{\mu\nu}$ having a signature $(+,-,-,-)$. The spin-1/2 fermion field is denoted
by $\psi$ and $\gamma^{\mu}$ are the standard Dirac matrices satisfying the Clifford algebra $\{\gamma^{\mu},\gamma^{\nu}\}=2\eta^{\mu\mu}\mathds{1}_4$
with the unit matrix $\mathds{1}_4$ in four-dimensional spinor space. The fermion mass is called $m_{\psi}$. The piece $\widehat{\Qs}$ involves
all Lorentz-violating contributions to an arbitrary mass dimension. It is convenient to decompose the latter into operators of different spin structure
using the Dirac bilinears, cf.~Eq.~(2) in \cite{Kostelecky:2013rta}.

Due to the reasons outlined in the introduction, having classical point-particle analogs of fields in the SME is mandatory. The procedure to obtain
the corresponding classical Lagrangians was developed in \cite{Kostelecky:2010hs}. In the center of the method there is a system of five polynomial
equations linking the four-momentum in the field-theory language to the four-velocity of the relativistic, pointlike particle in the classical regime.
These equations are given as follows:
\begin{subequations}
\label{eq:set-equations-lagrangians}
\begin{align}
\label{eq:dispersion-relation}
\mathcal{R}(p)&=0\,, \displaybreak[0]\\[2ex]
\label{eq:group-velocity-correspondence}
\frac{\mathrm{d}p_0}{\mathrm{d}p_i}&=-\frac{u^i}{u^0}\,,\quad i\in\{1,2,3\}\,, \displaybreak[0]\\[2ex]
\label{eq:euler-equation}
L&=-p_{\mu}u^{\mu}\,,\quad p_{\mu}=-\frac{\partial L}{\partial u^{\mu}}\,,
\end{align}
\end{subequations}
where $p_{\mu}$ is the four-momentum in the field-theory context and $u^{\mu}$ the four-velocity of the classical particle.
Equation (\ref{eq:dispersion-relation}) describes the dispersion equation resulting from the SME field theory. It involves the  four-momentum
components, the particle mass $m_{\psi}$, and Lorentz-violating coefficients. For the minimal SME the polynomial $\mathcal{R}(p)$ on its left-hand side is
quartic in $p_0$ at most. For the nonminimal SME the degree of this polynomial can be arbitrarily large, which makes it more involved to be solved.
Equation (\ref{eq:group-velocity-correspondence}) comprises three individual relationships that describe a correspondence between the group velocity
of the quantum-theoretical wave packet and the three-velocity of the pointlike particle. The minus sign on the right-hand side has to be taken into
account to respect the different positions of the spatial indices. Last but not least, \eqref{eq:euler-equation} is called Euler's formula. The latter
holds since the Lagrangian is supposed to be positively homogeneous of degree one in the velocity: $L(\lambda u)=\lambda L(u)$, $\lambda>0$. Positive
homogeneity is a very reasonable assumption since it makes the corresponding action parameterization-invariant, which should hold for any physical
system anyhow. We see that there is a direct correspondence between the four-momentum $p_{\mu}$ that is used in the field-theory language and the
canonical momentum computed via the first partial derivative of the Lagrangian, see \eqref{eq:euler-equation}. The latter involves a minus sign
to render the low-energy kinetic energy positive.

The set of equations stated depends on 9 dynamical quantities: the four-momentum components $p_{\mu}$ of the physical field, the four-velocity
components $u^{\mu}$ of the classical, relativistic particle, and the Lagrangian $L$. In principle four of these equations are needed to eliminate
the four-momentum and the remaining one allows for computing $L$. The most straightforward procedure would be to express the
four-momentum by the four-velocity via Eqs.~(\ref{eq:dispersion-relation}), (\ref{eq:group-velocity-correspondence}) and to insert the result
into \eqref{eq:euler-equation}. However experience shows that in practice this technique works only for very rare cases, e.g., certain isotropic
ones such as for the nonminimal coefficient $m^{(5)}_{00}$ \cite{Schreck:2014hga}. For the minimal cases studied in \cite{Kostelecky:2010hs} no
systematic procedure was used to solve the equations. However, since the latter are computationally more involved for the nonminimal SME it would
be desirable to have such a method available. Fortunately, sets of polynomial equations and systematic solution techniques have been studied in the
fields of algebra and algebraic geometry for several decades. The approach of dealing with Eqs.~(\ref{eq:set-equations-lagrangians}) will involve a
concept that is known as the Gr\"{o}bner basis.

\subsection{Gr\"{o}bner basis}

This section serves as a recapitulation of some basic algebraic terms and as an introduction to the notion of Gr\"{o}bner bases. Consider a
ring $(R,+,\cdot)$ with an addition + and a multiplication $\cdot$ where these operations can act on elements of the whole ring. An ideal $I$
is a subset of $R$ with the following properties:

\begin{itemize}

\item[1)] $(I,+)$ forms a subgroup of $(R,+)$,
\item[2)] for all $i\in I$ and $r\in R$ we have both $i\cdot r\in I$ and $r\cdot i\in I$.

\end{itemize}

That second property means that whenever an element of the group (regardless of whether it lies in the ideal or not) is multiplied with an
element of the ideal, the result will be in the ideal. For this reason, an ideal is capable of doing more than a group.
Now, a Gr\"{o}bner basis is a finite system of generating functions of an ideal in the polynomial ring $K[X_1\dots X_k]$ over the field $K$
where the $X_i$ can be interpreted as a set of $k$ variables. Consider an ideal $I$ that is generated by a set of $n$ polynomials $\{f_1\dots f_n\}$,
i.e., $I\equiv\langle f_1\dots f_n\rangle\subseteq K[X_1\dots X_k]$. An important algebraic problem is to decide whether a certain polynomial $g$ is an
element of the ideal, which means that it can be written as a linear combination $a_1f_1+\dots+a_nf_n$ where $a_1\dots a_n\in K[X_1 \dots X_k]$. In
general, it is cumbersome to take such a decision based on an arbitrary set of polynomials $\{f_1\dots f_n\}$ generating the ideal, i.e., the latter
basis may have been chosen badly to tackle this problem. The Gr\"{o}bner basis is much more suitable since it gives a simple criterion to check
whether $g\in I$.

The concept of a Gr\"{o}bner basis was introduced by Buchberger in his Ph.D. thesis \cite{Buchberger:1965} where some years later he named it after
his supervisor Gr\"{o}bner. At around the same time, Hironaka developed similar notions that he needed to proof his famous theorem for algebraic
varieties \cite{Hironaka:1964}. The Russian mathematician Gjunter had already thought about related ideas in the early 20th century and his papers
have been rediscovered recently, cf.~\cite{Rennschuh:2003} and references therein.

We do not want to delve into the mathematical theory of Gr\"{o}bner bases. For the scope of the paper it is important that finding a Gr\"{o}bner basis
allows us to solve a polynomial system of equations efficiently. From the perspective of mathematics, the solutions of such a system depend on the
generated ideal only but not on the generating functions chosen. Hence, replacing the (probably) unsuitable basis obtained from the equations, as they
stand, by a Gr\"{o}bner basis does not change the solution. Using the Gr\"{o}bner basis, the system can be solved step by step for each variable. The
usefulness of this approach will become evident in practice.

One of the most important of Buchberger's contributions was the algorithm that he developed to compute a Gr\"{o}bner basis. Nowadays this algorithm
or improved versions of it have already been implemented in various computer algebra systems. We will mostly work with \texttt{Mathematica} and
perform some cross checks with \texttt{Maple}. For the algorithm it is essential to introduce an ordering procedure for the individual monomials
that appear in the equations, based on an ordering of the variables such as $x_1>x_2>\dots$. There are several ordering conventions, e.g., the
lexicographic or the degree reverse lexicographic one. The lexicographic scheme was used by Buchberger in his Ph.D. thesis but it makes computations
quite tedious. Nevertheless it is the default ordering used by \texttt{Mathematica} and it suffices for our purposes.

\subsection{Example of the minimal SME}

As physical quantities are real numbers in general we have to consider the polynomial ring $\mathbb{R}[p_0,p_1,p_2,p_3,L]$. The coefficients of
polynomial equations that are usually investigated in mathematics are mere numbers. However, Eqs.~(\ref{eq:set-equations-lagrangians}) involve the
four-velocity components $u^{\mu}$, the particle mass $m_{\psi}$, and at least one Lorentz-violating coefficient, which are taken to be arbitrary
at first. Therefore, the Gr\"{o}bner bases will comprise these quantities as well, which increases the computing time. As a first example, a minimal
fermionic framework shall be considered that is described by \eqref{eq:action-fermionic-sme} with $\widehat{\Qs}|^{\widehat{a}^{(3)\mu}}=-a^{(3)\mu}\gamma_{\mu}$.
The corresponding field-theory operator is {\em CPT}-odd and of mass dimension 3. This makes the component coefficients $a^{(3)\mu}$ transform as an
observer four-vector with each component having mass dimension 1.

The corresponding dispersion equation involves the determinant of the left-hand side of the modified Dirac equation, cf.~for example Eq.~(31) in
\cite{Kostelecky:2013rta}. The resulting equation has both positive- and negative-energy solutions where the latter correspond to the energy of
antiparticles after the Feynman-St\"{u}ckelberg reinterpretation \cite{Kostelecky:2013rta,Schreck:2014qka}. For the special case under consideration,
the particle energy is well-known to take the following form:
\begin{equation}
\widetilde{E}^{(+)}-a_0^{(3)}=\sqrt{(\mathbf{p}-\mathbf{a}^{(3)})^2+m_{\psi}^2}\,,
\end{equation}
where $\mathbf{p}$ is the spatial momentum and $\mathbf{a}$ the spatial part of the vector coefficient $a_{\mu}^{(3)}$. Since in Minkowski spacetime these
component coefficients just lead to shifts in energy and spatial momentum the coefficients for a single particle species cannot be observed in
principle. Note that in the context of gravity the situation is different \cite{Kostelecky:2008in}, which is why the corresponding framework is still
interesting. Now, Eqs.~(\ref{eq:set-equations-lagrangians}) read as follows:
\begin{subequations}
\begin{align}
0&=p_0^2-2a_0^{(3)}p_0-p_1^2+2a_1^{(3)}p_1-p_2^2+2a_2^{(3)}p_2-p_3^2+2a_3^{(3)}p_3 \notag \\
&\phantom{{}={}}+(a_0^{(3)})^2-(\mathbf{a}^{(3)})^2-m_{\psi}^2\,, \displaybreak[0]\\[2ex]
0&=u^1p_0^2+u^0p_0p_1-(a_1^{(3)}u^0+a_0^{(3)}u^1)p_0\,, \displaybreak[0]\\[2ex]
0&=u^2p_0^2+u^0p_0p_2-(a_2^{(3)}u^0+a_0^{(3)}u^2)p_0\,, \displaybreak[0]\\[2ex]
0&=u^3p_0^2+u^0p_0p_3-(a_3^{(3)}u^0+a_0^{(3)}u^3)p_0\,, \displaybreak[0]\\[2ex]
0&=u^0p_0+u^1p_1+u^2p_2+u^3p_3+L\,.
\end{align}
\end{subequations}
Four of them have a degree of 2 whereas the final one has a degree of 1. With the algorithm implemented in \texttt{Mathematica}, the
Gr\"{o}bner basis can be computed readily. Thereby we follow two different possibilities. The first is to carry out the computation based on the
variable ordering $p_0>p_1>p_2>p_3>L$. This result involves 45 polynomials. Most of them are quite lengthy and they comprise all variables, which
is why they are not very helpful to obtain the solution of the system. However, there is one particular polynomial that involves the Lagrangian
only, which allows for computing $L$ immediately:
\begin{equation}
\label{eq:lagrangian-minimal-a-coefficients}
L^{\widehat{a}^{(3)\mu}\pm}=\pm m_{\psi}\sqrt{u^2}-a_{\mu}^{(3)}u^{\mu}\,.
\end{equation}
The second possibility investigated was the opposite variable ordering $L>p_3>p_2>p_1>p_0$. Here the calculation of the Gr\"{o}bner basis is much
faster and the result has 12 polynomials only that are much simpler. The difference to the first method is that among these polynomials there is
none that involves only the Lagrangian. However, a subset of the polynomials can be solved successively with respect to all the variables in analogy
to a linear system of equations that was brought into triangular form:
\begin{subequations}
\begin{align}
p_0&=a_0^{(3)}\pm \frac{m_{\psi}u^0}{\sqrt{u^2}}\,, \displaybreak[0]\\[2ex]
p_1&=-\frac{u^1}{u^0}p_0+a_1^{(3)}+\frac{a_0^{(3)}u^1}{u^0}\,, \displaybreak[0]\\[2ex]
p_2&=-\frac{u^2}{u^0}p_0+a_2^{(3)}+\frac{a_0^{(3)}u^2}{u^0}\,, \displaybreak[0]\\[2ex]
p_3&=-\frac{u^3}{u^0}p_0+a_3^{(3)}+\frac{a_0^{(3)}u^3}{u^0}\,, \displaybreak[0]\\[2ex]
L&=-(p_0u^0+p_1u^1+p_2u^2+p_3u^3)\,.
\end{align}
\end{subequations}
Inserting these equations into each other step by step leads to the same result as given in \eqref{eq:lagrangian-minimal-a-coefficients}, which
corresponds to the Lagrangian obtained in \cite{Kostelecky:2010hs}. The particle-antiparticle property of the fermionic field theory manifests
in two distinct Lagrangians having opposite signs before the first term. The resulting classical Lagrangian is related to what is called a Randers
structure in Euclidean geometry. We will come back to this point in more detail in \secref{sec:finsler-structures-properties}.

From the procedure intimated we learn several things. The variable ordering with $p_0$ the greatest and $L$ the smallest leads to an equation that
can be solved directly to obtain the Lagrangian. The drawbacks are that it takes long to obtain the Gr\"{o}bner basis and the latter involves
a large number of complicated polynomials that are not needed for our purpose. The reverse ordering of the variables with $L$ the greatest and
$p_0$ the smallest results in a system of equations that can be solved easily step by step to compute $L$ at the end. This ordering is preferable
as it is fast and does not lead to plenty of useless overhead. Furthermore, each of the equations can be simplified on its own, which will be
helpful for the nonminimal cases to be studied. Hence, we will work with the second ordering of variables. Other minimal frameworks such as the
case of the $b_{\mu}$ coefficients were tested successfully using the same algorithm, which opens the pathway to study the nonminimal SME.

\section{Classical Lagrangians for the nonminimal SME}
\label{sec:classical-lagrangians-nonminimal-sme}
\setcounter{equation}{0}

The computational technique outlined in the last section will now be applied to the nonminimal SME. The first Lagrangian for a nonminimal
fermionic framework was obtained in \cite{Schreck:2014hga} for the single nonvanishing coefficient $m^{(5)}_{00}$. The latter is associated to a
{\em CPT}-even Lorentz-violating operator of mass dimension 5. It was chosen because it leads to an isotropic dispersion relation and it was
thought to be the simplest case that could be investigated in the nonminimal SME fermion sector. The calculation was carried out nonperturbatively in
Lorentz violation resulting in a quite complicated Lagrangian whose properties were difficult to understand. An expansion of $L$ in $m^{(5)}_{00}$
at first order produced a much more illuminating result that was suitable to be used in a physics context. For example, the behavior of the
corresponding charged classical particle in electromagnetic fields was studied with this first-order Lagrangian.

For this reason, right at the start we will perform an expansion of the polynomial equations at first order in Lorentz violation. Note that
the validity of such an expansion in the nonminimal framework breaks down when the particle energy gets high enough, e.g., comparable to a
certain power of the inverse Lorentz-violating coefficients involved. This is not considered to be a problem since for classical physics to
be valid, the energy should not be too high anyhow, i.e., it should be much lower than the typical energies of a regime where quantum
effects will play a role.

\subsection{Five-dimensional part of the scalar operator \texorpdfstring{$\boldsymbol{\widehat{m}}$}{m-hat}}
\label{sec:lagrangian-operator-m5}

First of all, we are interested in the operator $\widehat{\Qs}|^{\widehat{m}}=-\widehat{m}\mathds{1}_4$. The latter transforms as an observer Lorentz scalar,
it is {\em CPT}-even, and it only exists in the nonminimal SME. Hence, there is no Lorentz-violating dimension-3 analog. It can be decomposed
into contributions of mass dimension $d$ according to
\begin{equation}
\widehat{m}=\sum_{d \text{ odd}} \widehat{m}^{(d)}\equiv \sum_{d \text{ odd}} m^{(d)\alpha_1\dots\alpha_{d-3}}p_{\alpha_1}\dots p_{\alpha_{d-3}}\,.
\end{equation}
In this paper we restrict ourselves to the lowest-dimensional contributions of the nonminimal SME, which in this case is the dimension-5 operator
$\widehat{m}^{(5)}\equiv m^{(5)\mu\nu}p_{\mu}p_{\nu}$ including the 16 component coefficients $m^{(5)}_{\mu\nu}$ that are of mass dimension
$-1$. The set of coefficients $m^{(5)}_{\mu\nu}$ can be regarded as a symmetric $(4\times 4)$ matrix, i.e., only 10 coefficients are independent
of each other. In analogy to \cite{Schreck:2014hga},
the component coefficients are put into distinct groups. The single coefficient $m^{(5)}_{00}$, which was the base of the studies carried out
in \cite{Schreck:2014hga}, will be called ``temporal'' as it involves two timelike indices. The set of three coefficients $m^{(5)}_{0i}$ with
$i\in\{1\dots 3\}$ are denoted as ``mixed'' and the set of six independent coefficients $m^{(5)}_{ij}$ for $i,j\in \{1\dots 3\}$ are named
``spatial.'' Such a classification of component coefficients has proven to be reasonable because the structure of the resulting equations
for the groups differ from each other. Similar decompositions will be performed for component coefficients of other Lorentz-violating operators
considered later.

First of all we intend to complement the results given in \cite{Schreck:2014hga}, i.e., the remaining component coefficients of $\widehat{m}^{(5)}$
will be taken into account. In the latter reference an observer frame was considered with $m^{(5)}_{00}$ being the only nonvanishing coefficient.
The first-order expansion of the corresponding Lagrangian is ready to be taken from Eq.~(6.1) in \cite{Schreck:2014hga}:
\begin{equation}
\label{eq:lagrangian-m5-temporal}
L^{\widehat{m}^{(5)}\pm}|_{\mathrm{temp}}=\pm m_{\psi}\Big[\sqrt{u^2}+\frac{m_{\psi}}{\sqrt{u^2}}m^{(5)}_{00}(u^0)^2+\dots\Big]\,,
\end{equation}
where the ellipses indicate higher-order contributions in $m_{\psi}m^{(5)}_{00}$.
Next, we consider an observer frame with the three nonvanishing mixed coefficients $m^{(5)}_{0i}$. At first order in Lorentz violation,
Eqs.~(\ref{eq:set-equations-lagrangians}) read as follows:
\begin{subequations}
\label{eq:set-equations-m5-mixed}
\begin{align}
0&=p_0^2+4m_{\psi}m^{(5)}_{01}p_0p_1+4m_{\psi}m^{(5)}_{02}p_0p_2+4m_{\psi}m^{(5)}_{03}p_0p_3-\mathbf{p}^2-m_{\psi}^2\,, \displaybreak[0]\\[2ex]
0&=(u^1-2m_{\psi}m^{(5)}_{01}u^0)p_0^2+(u^0+2m_{\psi}m^{(5)}_{01}u^1)p_0p_1 \notag \\
&\phantom{{}={}}+2m_{\psi}m^{(5)}_{02}u^1p_0p_2+2m_{\psi}m^{(5)}_{03}u^1p_0p_3\,, \displaybreak[0]\\[2ex]
0&=(u^2-2m_{\psi}m^{(5)}_{02}u^0)p_0^2+2m_{\psi}m^{(5)}_{01}u^2p_0p_1 \notag \\
&\phantom{{}={}}+(u^0+2m_{\psi}m^{(5)}_{02}u^2)p_0p_2+2m_{\psi}m^{(5)}_{03}u^2p_0p_3\,, \displaybreak[0]\\[2ex]
0&=(u^3-2m_{\psi}m^{(5)}_{03}u^0)p_0^2+2m_{\psi}m^{(5)}_{01}u^3p_0p_1 \notag \\
&\phantom{{}={}}+2m_{\psi}m^{(5)}_{02}u^3p_0p_2+(u^0+2m_{\psi}m^{(5)}_{03}u^3)p_0p_3\,, \displaybreak[0]\\[2ex]
0&=p_0u^0+p_1u^1+p_2u^2+p_3u^3+L\,.
\end{align}
\end{subequations}
Although the framework considered is nonminimal the individual equations are mostly quadratic since higher-order terms in Lorentz violation
have been omitted. The polynomials involve additional products of the field energy and the spatial momentum components, though. Now the
Gr\"{o}bner basis of the ideal generated by these polynomials can be computed. It comprises 66 polynomials but the computation time stays
within reasonable limits. We pick the useful polynomials of the basis that allow for a successive solution of the system. This procedure is
demonstrated in the first part of \appref{sec:obtaining-lagrangian-m5} in detail. At first order in Lorentz violation the resulting Lagrangian
reads
\begin{equation}
\label{eq:lagrangian-m5-mixed}
L^{\widehat{m}^{(5)}\pm}|_{\mathrm{mixed}}=\pm m_{\psi}\Big[\sqrt{u^2}+\frac{2m_{\psi}}{\sqrt{u^2}}\sum_i m^{(5)}_{0i}u^i+\dots\Big]\,,
\end{equation}
with the ellipses standing for higher-order terms of $m_{\psi}m^{(5)}_{0i}$. Now a similar computation can be carried out in an observer frame where
the six coefficients $m^{(5)}_{ij}$ involving only spatial indices are nonvanishing, cf.~the second part of \appref{sec:obtaining-lagrangian-m5}.
The computed Lagrangian is given by:
\begin{equation}
\label{eq:lagrangian-m5-spatial}
L^{\widehat{m}^{(5)}\pm}|_{\mathrm{spatial}}=\pm m_{\psi}\Big[\sqrt{u^2}+\frac{m_{\psi}}{\sqrt{u^2}}\sum_{i,j} m^{(5)}_{ij}u^iu^j+\dots\Big]\,.
\end{equation}
Comparing Eqs.~(\ref{eq:lagrangian-m5-temporal}) -- (\ref{eq:lagrangian-m5-spatial}) with each other we see that these can be generalized to an
arbitrary observer frame involving the whole set of 10 independent coefficients $m^{(5)}_{\mu\nu}$. The result can be cast into the form
\begin{equation}
\label{eq:lagrangian-m5-complete}
L^{\widehat{m}^{(5)}\pm}=\pm m_{\psi}\Big[\sqrt{u^2}+\frac{m_{\psi}\widehat{m}^{(5)}_{\ast}}{\sqrt{u^2}}+\dots\Big]\,,\quad \widehat{m}^{(5)}_{\ast}\equiv m^{(5)}_{\mu\nu}u^{\mu}u^{\nu}\,.
\end{equation}
Note the asterisk that has been introduced to distinguish $m^{(5)}_{\mu\nu}u^{\mu}u^{\nu}$ from the corresponding combination in momentum
space, i.e., $\widehat{m}^{(5)}\equiv m^{(5)\mu\nu}p_{\mu}p_{\nu}$.
A few remarks are in order. First, there are two Lagrangians differing by a global sign. The plus sign can be understood to be associated to
particles where the minus sign is associated to antiparticles in the corresponding quantum theory. Second, for vanishing Lorentz-violating
component coefficients, the standard result $L^{\pm}=\pm m_{\psi}\sqrt{u^2}$ is recovered. Third, the Lorentz-violating contribution involves
the dimensionless product of the particle mass and the Lorentz-violating coefficients. Furthermore, the velocity dependence of the novel
contribution differs from that of the standard result. Fourth, the Lorentz-violating part can be written in a very compact form where the
recently introduced symbol $\widehat{m}^{(5)}_{\ast}$ is the mirror image of $\widehat{m}^{(5)}$ in momentum space. To obtain the former
from the latter all four-momenta are replaced by four-velocities and the index positions are adapted. Fifth, for the Lorentz-violating
part to be still perturbative, $u^2$ should be far enough away from zero. In the proper-time parameterization this means that the magnitude
of the particle velocity is supposed to be small enough compared to the speed of light. This is the analog of the claim for particle momenta
and energies to be much smaller than certain inverse powers of SME coefficients.

As a crosscheck for \eqref{eq:lagrangian-m5-complete}, the associated canonical momentum can be computed according to
$p_{\mu}=-\partial L/\partial u^{\mu}$. It has to fulfill Eqs.~(\ref{eq:set-equations-lagrangians}) for the full set of 10 independent
component coefficients $m^{(5)\mu\nu}$ at first order in these coefficients. Performing this computation numerically with our computer
algebra system chosen shows that this is the case.

\subsection{Leading nonminimal parts of the operators \texorpdfstring{$\boldsymbol{\widehat{a}}$}{a-hat}, \texorpdfstring{$\boldsymbol{\widehat{c}}$}{c-hat}, \texorpdfstring{$\boldsymbol{\widehat{e}}$}{e-hat}, and \texorpdfstring{$\boldsymbol{\widehat{f}}$}{f-hat}}

The method that we use for computing classical Lagrangians of the nonminimal SME fermion sector was demonstrated in the last section in detail,
based on the dimension-5 part of the scalar operator $\widehat{m}$. There are other interesting operators whose classical-particle equivalents
will be obtained in the current section. This concerns the leading nonminimal contributions of the vectors $\widehat{a}^{\mu}$, $\widehat{c}^{\,\mu}$,
the scalar $\widehat{e}$, and the pseudoscalar $\widehat{f}$, cf.~Table I in \cite{Kostelecky:2013rta} for a summary of their properties. The
dimensional decompositions for these operators read as follows:
\begin{subequations}
\label{eq:dimensional-expansions-various-operators}
\begin{align}
\widehat{a}^{\mu}&=\sum_{d \text{ odd}} \widehat{a}^{(d)\mu}\equiv \sum_{d \text{ odd}} a^{(d)\mu\alpha_1\dots\alpha_{d-3}}p_{\alpha_1}\dots p_{\alpha_{d-3}}\,, \displaybreak[0]\\[2ex]
\widehat{c}^{\,\mu}&=\sum_{d \text{ even}} \widehat{c}^{\,(d)\mu}\equiv \sum_{d \text{ even}} c^{(d)\mu\alpha_1\dots\alpha_{d-3}}p_{\alpha_1}\dots p_{\alpha_{d-3}}\,, \displaybreak[0]\\[2ex]
\widehat{e}&=\sum_{d \text{ even}} \widehat{e}^{\,(d)}\equiv \sum_{d \text{ even}} e^{(d)\alpha_1\dots \alpha_{d-3}}p_{\alpha_1}\dots p_{\alpha_{d-3}}\,, \displaybreak[0]\\[2ex]
\widehat{f}&=\sum_{d \text{ even}} \widehat{f}^{\,(d)}\equiv \sum_{d \text{ even}} f^{(d)\alpha_1\dots \alpha_{d-3}}p_{\alpha_1}\dots p_{\alpha_{d-3}}\,.
\end{align}
\end{subequations}
As before, we consider field theories based on the action of \eqref{eq:action-fermionic-sme} for four different choices of the Lorentz-violating
part $\widehat{\Qs}$:
\begin{subequations}
\begin{align}
\widehat{\Qs}|^{\widehat{a}^{(5)\mu}}&=-\widehat{a}^{(5)\mu}\gamma_{\mu}=-a^{(5)\mu\nu\varrho}\gamma_{\mu}p_{\nu}p_{\varrho}\,, \displaybreak[0]\\[2ex]
\widehat{\Qs}|^{\widehat{c}^{\,(6)\mu}}&=\widehat{c}^{\,(6)\mu}\gamma_{\mu}=c^{(6)\mu\nu\varrho\sigma}\gamma_{\mu}p_{\nu}p_{\varrho}p_{\sigma}\,, \displaybreak[0]\\[2ex]
\widehat{\Qs}|^{\widehat{e}^{\,(6)}}&=\widehat{e}^{\,(6)}\mathds{1}_4=e^{(6)\mu\nu\varrho}p_{\mu}p_{\nu}p_{\varrho}\mathds{1}_4\,, \displaybreak[0]\\[2ex]
\widehat{\Qs}|^{\widehat{f}^{\,(6)}}&=\mathrm{i}\widehat{f}^{\,(6)}\gamma_5=\mathrm{i}f^{(6)\mu\nu\varrho}p_{\mu}p_{\nu}p_{\varrho}\gamma_5\,,
\end{align}
\end{subequations}
where $\gamma_5\equiv \mathrm{i}\gamma^0\gamma^1\gamma^2\gamma^3$.
The computations are completely analogous to what we performed for $\widehat{m}^{(5)}$ in \secref{sec:lagrangian-operator-m5}. The only difference
is that they require more resources and time. This is especially the case for $\widehat{c}^{\,(6)\mu}$ and $\widehat{f}^{\,(6)}$.
At first order in Lorentz violation the Lagrangians obtained from the Gr\"{o}bner bases can be brought into the following form:
\begin{subequations}
\label{eq:lagrangians-nonminimal-a5-c6-e6-f6}
\begin{align}
\label{eq:lagrangian-a5}
L^{\widehat{a}^{(5)\mu}\pm}&=m_{\psi}\Big[\pm\sqrt{u^2}-\frac{m_{\psi}\widehat{a}^{(5)}_{\ast}}{u^2}+\dots\Big]\,,\quad \widehat{a}^{(5)}_{\ast}\equiv a^{(5)}_{\mu\nu\varrho}u^{\mu}u^{\nu}u^{\varrho}\,, \displaybreak[0]\\[2ex]
\label{eq:lagrangian-c6}
L^{\widehat{c}^{\,(6)\mu}\pm}&=\pm m_{\psi}\Big[\sqrt{u^2}-\frac{m_{\psi}^2\widehat{c}^{\,(6)}_{\ast}}{(u^2)^{3/2}}+\dots\Big]\,,\quad \widehat{c}^{\,(6)}_{\ast}\equiv c^{(6)}_{\mu\nu\varrho\sigma}u^{\mu}u^{\nu}u^{\varrho}u^{\sigma}\,, \displaybreak[0]\\[2ex]
\label{eq:lagrangian-e6}
L^{\widehat{e}^{\,(6)}\pm}&=m_{\psi}\Big[\pm\sqrt{u^2}+\frac{m_{\psi}^2\widehat{e}^{\,(6)}_{\ast}}{u^2}+\dots\Big]\,,\quad \widehat{e}^{\,(6)}_{\ast}\equiv e^{(6)}_{\mu\nu\varrho}u^{\mu}u^{\nu}u^{\varrho}\,, \displaybreak[0]\\[2ex]
\label{eq:lagrangian-f6}
L^{\widehat{f}^{\,(6)}\pm}&=\pm m_{\psi}\Big[\sqrt{u^2}+\frac{m_{\psi}^4\big(\widehat{f}^{\,(6)}_{\ast}\big)^2}{2(u^2)^{5/2}}+\dots\Big]\,,\quad \widehat{f}^{\,(6)}_{\ast}\equiv f^{(6)}_{\mu\nu\varrho}u^{\mu}u^{\nu}u^{\varrho}\,.
\end{align}
\end{subequations}
Using $p_{\mu}=-\partial L/\partial u^{\mu}$, these results were checked numerically to fulfill Eqs.~(\ref{eq:set-equations-lagrangians})
at first order in Lorentz violation.
Several remarks are in order. First, all Lagrangians reduce to the standard result $L^{\pm}=\pm m_{\psi}\sqrt{u^2}$ for vanishing component
coefficients. Second, due to observer Lorentz invariance, all relativistic Lagrangians depend on Lorentz scalars only such as $u^2$ or
total contractions of the component coefficients with four-velocities. The latter are analogs of the Lorentz-violating operators
in momentum space that appear in the dimensional expansion of \eqref{eq:dimensional-expansions-various-operators}. The dimensions
of the Lorentz-violating contributions with respect to the velocity are consistent with the standard term. For example, in
\eqref{eq:lagrangian-a5} the component coefficients are contracted with three velocity four-vectors, which is why there is no choice
other than $u^2$ in the denominator. Third, since a Lagrangian is of mass dimension 1 the Lorentz-violating contributions must involve
additional powers of the fermion mass. Fourth, the two distinct signs, which are connected to the particle-antiparticle solutions in field
theory, appear globally for the Lagrangians of the $\widehat{m}$, $\widehat{c}$ and $\widehat{f}$ operator whereas they appear before the standard term
only in case of the $\widehat{a}$ and $\widehat{e}$ operators. Fifth, according to Eq.~(27) in \cite{Kostelecky:2013rta}, several operators
are related to each other. This explains the similarities between \eqref{eq:lagrangian-a5}, \eqref{eq:lagrangian-e6} and \eqref{eq:lagrangian-m5-complete},
\eqref{eq:lagrangian-c6}. We will come back to this point in more detail below. Sixth, it is known that the minimal $\widehat{c}$ and $\widehat{f}$ operators are linked by the
perturbative relationship $c_{\mu\nu}^{(4)}\approx-f^{(4)}_{\mu}f^{(4)}_{\nu}/2$ \cite{Altschul:2006ts}. Comparing \eqref{eq:lagrangian-c6} to \eqref{eq:lagrangian-f6}
it is evident that a similar connection holds for $\widehat{c}^{\,(6)}_{\ast}$ and $\widehat{f}^{\,(6)}_{\ast}$, at least at
first order in Lorentz violation:
\begin{equation}
\widehat{c}^{\,(6)}_{\ast}=-\frac{1}{2}\big(\widehat{f}^{\,(6)}_{\ast}\big)^2+\dots\,.
\end{equation}
Because of dimensional consistency, the powers of $m_{\psi}$ and $u^2$ have to be adapted in the corresponding terms of the Lagrangians.
Last but not least, in \cite{Girelli:2006fw} a classical Lagrangian was obtained for a particular fermion dispersion relation modified
by a nonminimal operator, cf.~the two-dimensional restriction given by their Eq.~(22). The form of the dispersion law suggests that this
is a special case of $\widehat{a}^{(5)\mu}$ with the identification $a^{(5)}_{111}=-\alpha/(2M)$ and all remaining component coefficients
set to zero. Here $\alpha$ is a dimensionless parameter and $M$ the scale related to new physics modifying the dispersion relation
(presumably the Planck scale).
The corresponding classical Lagrangian is given by their Eq.~(28). Here it must be taken into account that their convention differs
from ours by a global minus sign of the Lagrangian. Hence, their result can be found as a special case of the second Lagrangian given
in \eqref{eq:lagrangian-a5} when the global sign is adapted.

In \cite{Kostelecky:2010hs} various relativistic point-particle Lagrangians of the minimal SME were obtained. For our purpose, their
Eqs.~(8), (9), and (10) are essential giving the Lagrangians for the minimal $\widehat{a}$, $\widehat{c}$, $\widehat{e}$, and
$\widehat{f}$ operators. A first-order expansion in Lorentz violation leads to the results
\begin{subequations}
\begin{align}
\label{eq:lagrangian-a3}
L^{\widehat{a}^{(3)\mu}\pm}&=\pm m_{\psi}\sqrt{u^2}-\widehat{a}^{\,(3)}_{\ast}\,,\quad \widehat{a}^{\,(3)}_{\ast}\equiv a^{(3)}_{\mu}u^{\mu}\,, \displaybreak[0]\\[2ex]
L^{\widehat{c}^{\,(4)\mu}}&=-m_{\psi}\Big[\sqrt{u^2}-\frac{\widehat{c}^{\,(4)}_{\ast}}{\sqrt{u^2}}+\dots\Big]\,,\quad \widehat{c}^{\,(4)}_{\ast}\equiv c^{(4)}_{\mu\nu}u^{\mu}u^{\nu}\,, \displaybreak[0]\\[2ex]
L^{\widehat{e}^{\,(4)}}&=m_{\psi}\Big[-\sqrt{u^2}+\widehat{e}^{\,(4)}_{\ast}+\dots\Big]\,,\quad \widehat{e}^{\,(4)}_{\ast}\equiv e^{(4)}_{\mu}u^{\mu}\,, \displaybreak[0]\\[2ex]
L^{\widehat{f}^{\,(4)}}&=-m_{\psi}\Big[\sqrt{u^2}+\frac{(\widehat{f}^{\,(4)}_{\ast})^2}{2\sqrt{u^2}}+\dots\Big]\,,\quad \widehat{f}^{\,(4)}_{\ast}\equiv f^{(4)}_{\mu}u^{\mu}\,,
\end{align}
\end{subequations}
where for the last three only one of the two possible signs is given. Note that the first of these results was also obtained in
\eqref{eq:lagrangian-minimal-a-coefficients} to demonstrate the method of Gr\"{o}bner basis for a simple example. The latter Lagrangian
does not have higher-order terms in Lorentz violation. Comparing these minimal results to their nonminimal analogs of
\eqref{eq:lagrangians-nonminimal-a5-c6-e6-f6} reveals a very similar structure.

Another excellent crosscheck for the results is that four of the Lagrangians can be reduced to two by resorting to the effective
coefficients in the SME fermion sector. Using field redefinitions in the effective field-theory context it was shown that certain
sets of component coefficients combine to form observables. The latter combinations are called effective coefficients. For us the first two of
Eq.~(27) in \cite{Kostelecky:2013rta} are important. This concerns the operators $\widehat{e}^{\,(4)}$,
$\widehat{a}^{(5)}$ and $\widehat{m}^{(5)}$, $\widehat{c}^{\,(6)}$. The resulting Lagrangians are expressed in terms of the effective
$a$ and $c$ coefficients as follows:
\begin{subequations}
\begin{align}
L^{\substack{\widehat{e}^{\,(4)} \\ \widehat{a}^{(5)}}}&\equiv m_{\psi}\bigg[\pm\sqrt{u^2}-\frac{m_{\psi}\widehat{a}^{(5)}_{\ast}}{u^2}+\widehat{e}^{\,(4)}_{\ast}+\dots\bigg] \notag \\
&=m_{\psi}\bigg[\pm\sqrt{u^2}-\frac{m_{\psi}\widehat{a}^{(5)}_{\ast}}{u^2}+\frac{\eta_{\mu\nu}u^{\mu}u^{\nu}\widehat{e}^{\,(4)}_{\ast}}{u^2}+\dots\bigg] \notag \\
&=m_{\psi}\bigg[\pm\sqrt{u^2}-\frac{m_{\psi}\widehat{a}_{\mathrm{eff,\ast}}^{(5)}}{u^2}+\dots\bigg]\equiv L^{\widehat{a}_{\mathrm{eff}}^{(5)}}\,,
\end{align}
and
\begin{align}
L^{\substack{\widehat{m}^{(5)} \\ \widehat{c}^{\,(6)}}}&\equiv\pm m_{\psi}\bigg[\sqrt{u^2}-\frac{m_{\psi}^2\widehat{c}^{\,(6)}_{\ast}}{(u^2)^{3/2}}+\frac{m_{\psi}\widehat{m}^{(5)}_{\ast}}{\sqrt{u^2}}+\dots\bigg] \notag \\
&=\pm m_{\psi}\bigg[\sqrt{u^2}-\frac{m_{\psi}^2\widehat{c}^{\,(6)}_{\ast}}{(u^2)^{3/2}}+\frac{m_{\psi}\eta_{\mu\nu}u^{\mu}u^{\nu}\widehat{m}^{(5)}_{\ast}}{(u^2)^{3/2}}+\dots\bigg] \notag \\
&=\pm m_{\psi}\bigg[\sqrt{u^2}-\frac{m_{\psi}^2\widehat{c}^{\,(6)}_{\mathrm{eff},\ast}}{(u^2)^{3/2}}+\dots\bigg]\equiv L^{\widehat{c}_{\mathrm{eff}}^{\,(6)}}\,,
\end{align}
with
\begin{equation}
\widehat{a}_{\mathrm{eff,\ast}}^{(5)}\equiv (a^{(5)}_{\mathrm{eff}})_{\mu\nu\varrho}u^{\mu}u^{\nu}u^{\varrho}\,,\quad \widehat{c}_{\mathrm{eff,\ast}}^{\,(6)}\equiv (c^{(6)}_{\mathrm{eff}})_{\mu\nu\varrho\sigma}u^{\mu}u^{\nu}u^{\varrho}u^{\sigma}\,.
\end{equation}
\end{subequations}
In the definition of the effective coefficients, component coefficients being part of different operators have different mass dimension.
According to the first of Eq.~(27) in \cite{Kostelecky:2013rta} the operator $\widehat{e}^{\,(6)}$ is linked to the operator
$\widehat{a}^{(7)\mu}$. Since we did not derive the classical Lagrangian based on $\widehat{a}^{(7)\mu}$ we cannot express
\eqref{eq:lagrangian-e6} via effective coefficients. The Lagrangian based on $\widehat{f}^{(6)}$, \eqref{eq:lagrangian-f6}, plays
a special role as well. The operator $\widehat{f}$ is known not to contribute to observables at linear order in Lorentz violation.
This is why effective coefficients comprising components of $\widehat{f}$ were not introduced in \cite{Kostelecky:2013rta} restricting
the analysis mainly to linear order in Lorentz violation, which is difficult enough.

\subsection{Isotropic parts}
\label{sec:isotropic-parts}

In the coming section, the classical Lagrangians derived will serve as the base for phenomenological studies.
To establish a connection with experiment we choose convenient sets of coefficients, which are the isotropic ones, cf.~Eqs.~(97), (98)
in \cite{Kostelecky:2013rta}. In suitable observer frames, the Lagrangians of \eqref{eq:lagrangians-nonminimal-a5-c6-e6-f6} read
(see \appref{sec:obtaining-isotropic-lagrangians} for calculational details):
\begin{subequations}
\begin{align}
\ring{L}{}^{\widehat{a}_{\mathrm{eff}}^{(5)}}&=-m_{\psi}\left\{\sqrt{u^2}+m_{\psi}\left[\ring{a}{}^{(5)}_0(u^0)^2+\ring{a}{}^{(5)}_2\mathbf{u}^2\right]\frac{u^0}{u^2}\right\}\,, \\[2ex]
\ring{L}{}^{\widehat{c}_{\mathrm{eff}}^{\,(6)}}&=m_{\psi}\left\{-\sqrt{u^2}+m_{\psi}^2\left[\ring{c}{}^{(6)}_0(u^0)^4+\ring{c}{}^{(6)}_2(u^0)^2\mathbf{u}^2+\ring{c}{}^{(6)}_4\mathbf{u}^4\right]\frac{1}{(u^2)^{3/2}}\right\}\,, \\[2ex]
\ring{L}{}^{\widehat{e}^{\,(6)}}&=m_{\psi}\left\{-\sqrt{u^2}+m_{\psi}^2\left[\ring{e}{}^{(6)}_0(u^0)^2+\ring{e}{}^{(6)}_2\mathbf{u}^2\right]\frac{u^0}{u^2}\right\}\,, \\[2ex]
\ring{L}{}^{\widehat{f}^{\,(6)}}&=-m_{\psi}\left\{\sqrt{u^2}+m_{\psi}^4\left[\ring{f}{}^{(6)}_0(u^0)^2+\ring{f}{}^{(6)}_2\mathbf{u}^2\right]^2\frac{(u^0)^2}{2(u^2)^{5/2}}\right\}\,.
\end{align}
\end{subequations}
The isotropic contributions are denoted by a ring diacritic, which is a standard notation in this context. The Lagrangians with
a minus sign before the standard term are given as these will be needed in what follows. The isotropic Lagrangians depend on
$u^0$ and $|\mathbf{u}|$ only, as expected. We want to remind the reader that the isotropic coefficients for $\widehat{f}$ and
$\widehat{e}$ alone (with the latter not being comprised in the effective operator $\widehat{a}_{\mathrm{eff}}$) were not introduced
in \cite{Kostelecky:2013rta}. However, the index structure of both types of component coefficients is the same as for the operator
$\widehat{a}^{(5)\mu}$. Hence, we had to introduce
\begin{subequations}
\begin{align}
\ring{e}{}_0^{(6)}&\equiv e^{(6)}_{000}\,,\quad \ring{e}_2^{(6)}\equiv e^{(6)}_{0jj}\,, \\[2ex]
\ring{f}{}_0^{(6)}&\equiv f^{(6)}_{000}\,,\quad \ring{f}_2^{(6)}\equiv f^{(6)}_{0jj}\,.
\end{align}
\end{subequations}

\section{Experimental sensitivity on Lorentz-violating coefficients}
\label{sec:experimental-sensitivity}
\setcounter{equation}{0}

In the current section, we will obtain experimental sensitivities on a subset of nonminimal SME coefficients based on the Lagrangians
previously derived. Since these Lagrangians were obtained to do classical physics based on the field-theory concept of the SME we
have to consider kinematic experiments. Reliable tests of particle kinematics for velocities approaching the speed of light were
carried out in the first and early second half
of the 20th century. Later on, Special Relativity was commonly accepted to be the correct theory describing nature, which was why
these classical kinematic tests ceased to be made. Due to an active search for {\em CPT}- and Lorentz violation over the recent
decades, tests of Special Relativity have had their revival. However, the novel experiments performed rely on different techniques
such as sophisticated optical setups or ultra-high energetic cosmic particles. In accelerator and collider physics, the laws of
Special Relativity are mostly used as an input instead of testing them directly.

So far, all Lagrangians have been given for arbitrary parameterizations of particle trajectories. As from now, particle worldlines
will be parameterized by proper time. For this choice we have to set $u^0=c$ and $\mathbf{u}=\mathbf{v}$ where $c$ is the (reinstated)
speed of light and $\mathbf{v}$ is the three-velocity of the particle. Introducing the Lorentz factor $\gamma\equiv 1/\sqrt{1-\beta^2}$
with $\beta\equiv |\mathbf{v}|/c\equiv v/c$ (where possible confusions with the Dirac matrices shall be avoided) one obtains:
\begin{align}
\label{eq:kinematic-lagrangians}
\ring{L}{}^{\widehat{\mathcal{o}}}&=m_{\psi}c^2\left\{-\frac{1}{\gamma}+(m_{\psi}c^2)^{r_1^{(i)}}r_2^{(i)}\Big[\ring{i}_0+\ring{i}_2v^2+\ring{i}_4v^4\Big]^{r_3^{(i)}}\gamma^{r_4^{(i)}}\right\}\,.
\end{align}
For the five operators $\widehat{m}$, $\widehat{a}^{\mu}$, $\widehat{c}^{\,\mu}$, $\widehat{e}$, and $\widehat{f}$ considered,
the Lagrangian has been written in a generic form. Here $\widehat{\mathcal{o}}$ stands for one of these operators. What has to
be inserted for the wildcard character $i$ and the variables $r_k^{(i)}$ can be found in \tabref{tab:numbers-lagrangians}. These
Lagrangians describe the kinematics of a free, classical particle subject to Lorentz violation based on $\widehat{\mathcal{o}}$.

\begin{table}[t]
\centering
\begin{tabular}{ccccccc}
\toprule
Operator & $(i)$                                     & $r_1^{(i)}$ & $r_2^{(i)}$ & $r_3^{(i)}$ & $r_4^{(i)}$ \\
\colrule
$\widehat{a}^{(5)\mu}_{\mathrm{eff}}$   & $a^{(5)\mu}$ & 1 & $-1$   & 1 & 2 \\
$\widehat{c}^{\,(6)\mu}_{\mathrm{eff}}$ & $c^{\,(6)\mu}$ & 2 & 1      & 1 & 3 \\
$\widehat{e}^{\,(6)}$                   & $e^{\,(6)}$ & 2 & 1      & 1 & 2 \\
$\widehat{f}^{(6)}$                     & $f^{(6)}$ & 4 & $-1/2$ & 2 & 5 \\
\botrule
\end{tabular}
\caption{Parameters for the generic Lagrangian of \eqref{eq:kinematic-lagrangians} for each of the operators in the first column.}
\label{tab:numbers-lagrangians}
\end{table}

\subsection{Kinematic tests of Lorentz invariance}

In the kinematic tests of Special Relativity to be considered below, charged particles such as electrons and protons were set up
to propagate through both electric and magnetic fields. Hence, it is necessary to couple particles based on the free
Lagrangian of \eqref{eq:kinematic-lagrangians} to an external electromagnetic field with four-potential $(A^{\mu})=(\phi,\mathbf{A})$.
Here $\phi$ is the scalar potential and $\mathbf{A}$ the vector potential. To do so, the classical point-particle previously
considered is now assigned an electric charge $q$ (cf.~also \cite{Schreck:2014ama,Schreck:2014hga}). The physical situation is
described by a Lagrangian of the form
\begin{equation}
L_{\mathrm{em}}=\ring{L}{}^{\widehat{\mathcal{o}}}+q\mathbf{v}\cdot\mathbf{A}-q\phi\,.
\end{equation}
The corresponding Euler-Lagrange equations are then given by
\begin{subequations}
\begin{align}
\frac{\mathrm{d}}{\mathrm{d}t}\frac{\partial L_{\mathrm{em}}}{\partial\mathbf{v}}&=\frac{\partial L_{\mathrm{em}}}{\partial\mathbf{x}}\,, \\[2ex]
\label{eq:equations-of-motion}
\frac{\mathrm{d}}{\mathrm{d}t}\frac{\partial \ring{L}{}^{\widehat{\mathcal{o}}}}{\partial\mathbf{v}}&=\frac{\mathrm{d}\mathbf{p}}{\mathrm{d}t}=\frac{\mathrm{d}}{\mathrm{d}t}\Big(\frac{\partial \ring{L}{}^{\widehat{\mathcal{o}}}}{\partial v}\frac{\mathbf{v}}{v}\Big)=q\mathbf{v}\times\mathbf{B}+q\mathbf{E}\,,
\end{align}
\end{subequations}
where $\mathbf{p}$ is the particle three-momentum.
Consider a magnetic field pointing along the third axis of the coordinate system: $\mathbf{B}=B\widehat{\mathbf{e}}_z$. The magnitude of
the momentum does not change, which is why we obtain
\begin{equation}
\frac{1}{v}\frac{\partial \ring{L}{}^{\widehat{\mathcal{o}}}}{\partial v}\dot{\mathbf{v}}=qB\,\mathbf{v}\times\widehat{\mathbf{e}}_z\,.
\end{equation}
This equation is solved by a circular trajectory where the following connection holds between its radius $R^{(B)}$ and the remaining quantities:
\begin{equation}
\label{eq:relation-magnetic-field}
qBR^{(B)}=-\frac{\partial \ring{L}{}^{\widehat{\mathcal{o}}}}{\partial v}=-p\,,
\end{equation}
with the magnitude $p$ of the three-momentum. Similarly, in a radial electric field $\mathbf{E}=E\,\widehat{\mathbf{e}}_r$, a charged
particle moves on a circle as well since the electric field generates a central force. Then the magnitude of the particle momentum stays
constant again and the equations of motion are:
\begin{equation}
\frac{1}{v}\frac{\partial \ring{L}{}^{\widehat{\mathcal{o}}}}{\partial v}\dot{\mathbf{v}}=qE\,\widehat{\mathbf{e}}_r
\end{equation}
In analogy to the case of a homogeneous magnetic field, the radius $R^{(E)}$ of the trajectory is connected with the electric field strength
$E$ and the remaining physical quantities by
\begin{equation}
\label{eq:relation-electric-field}
qER^{(E)}=-v\frac{\partial \ring{L}{}^{\widehat{\mathcal{o}}}}{\partial v}=-pv=-\frac{p^2}{m_{\psi}(v)}\,,\quad m_{\psi}(v)\equiv\frac{p}{v}=\frac{1}{v}\frac{\partial \ring{L}{}^{\widehat{\mathcal{o}}}}{\partial v}\,,
\end{equation}
where we have introduced the modified relativistic mass $m_{\psi}(v)$.
There is some important information contained in Eqs.~(\ref{eq:relation-magnetic-field}) and (\ref{eq:relation-electric-field}). Quantities
that can be measured (the field strengths and the trajectory radius) or quantities known to a high level of accuracy (particle charge) can be
found on the left-hand sides of these equations. The right-hand sides comprise the particle dynamics affected by Lorentz violation. The
ultimate goal is to detect nonvanishing Lorentz-violating coefficients that modify the particle dynamics.

Due to the reasons mentioned at the beginning of the current section, we will have to rely on early papers to obtain experimental
sensitivities. Although there is a lot of potential in astrophysical experiments, our preference is on laboratory tests. The
latter run in a controlled environment and there are less assumptions that must be used as an input, which leads to more conservative
constraints.
Our first choice are experiments of the type performed in \cite{Rogers:1940}, \cite{Staub:1963}. The first of these
papers describes the measurement of electron velocities and masses for three different $\upbeta$-decay lines of Radium. To do so,
electrons are guided through a homogeneous magnetic and a cylindrical electric field where the measurements in the magnetic field
had been done elsewhere. The paper focuses on the propagation of electrons through a $90^{\circ}$ segment of a cylindrical capacitor
to obtain both the velocities and masses of electrons according to Eqs.~(\ref{eq:relation-magnetic-field}) and (\ref{eq:relation-electric-field}).
The experimenters put much rigor into keeping their setup stable, especially the distance between the capacitor plates. Therefore, they
are able to demonstrate the validity of Einstein's formula for mass increase as a function of velocity to an accuracy of about 1\%.

In \cite{Staub:1963} an apparatus is used to guide electrons and protons through a $180^{\circ}$ bending magnet and a $90^{\circ}$
segment of a cylindrical capacitor. First, electrons are sent through the apparatus with the magnetic field fixed to a
suitable value and the electric field varied such that the electrons hit the detector. Thereby the electron momentum is obtained
with \eqref{eq:relation-magnetic-field}. The corresponding magnetic field value $B_{\mathrm{e}}$, which will be needed later, is measured with
the proton resonance method. Now it is possible to compute the electron mass from the electric field strength according to
\eqref{eq:relation-electric-field}. However, since the latter is difficult to measure an approach is followed different from
\cite{Rogers:1940}. The whole experiment is repeated for protons, i.e., protons are used as a reference. Upon sending protons
through the apparatus the electric field is kept fixed and the magnetic field is adjusted until the protons are detected. With
this being the case, one gains knowledge of the proton momentum by measuring the magnetic field $B_{\mathrm{p}}$ with the proton resonance
method again. Then the electron mass is computed from \eqref{eq:relation-electric-field} with the electric field eliminated.
Since protons are much more massive than electrons they are nonrelativistic and it suffices to consider relativistic corrections
to their mass at leading order. The measurement technique is sophisticated and various error studies are performed. This leads
to a result that is a factor of around 20 better in comparison to \cite{Rogers:1940}.

\subsection{Constraints on Lorentz violation}

To obtain experimental sensitivities based on tests of relativistic kinematics described before, the modified particle momentum is
needed. The latter is obtained directly from the Lagrangian. The result can be written more conveniently in terms of powers of the
Lorentz factor where the related parameters $t_k^{(i)}$ are listed in \tabref{tab:parameters-momentum}:
\begin{equation}
\label{eq:explicit-result-momentum}
\ring{p}{}^{\widehat{\mathcal{o}}}=\frac{\partial \ring{L}{}^{\widehat{\mathcal{o}}}}{\partial v}=\gamma m_{\psi}v\bigg(1+(m_{\psi}c^2)^{\alpha^{(i)}}\sum_{k=0}^6 s_k^{(i)}\gamma^k\bigg)\,.
\end{equation}
A very useful dimensionless quantity, which was introduced in \cite{Rogers:1940}, is
\begin{subequations}
\begin{equation}
Y\equiv\frac{m_{\psi}(v)/m_{\psi}}{\sqrt{1+p^2/(m_{\psi}^2c^2)}}=\frac{p/(m_{\psi}v)}{\sqrt{1+p^2/(m_{\psi}^2c^2)}}\,.
\end{equation}
For the special Lagrangians considered it reads as follows:
\begin{equation}
\label{eq:quantity-y}
\ring{Y}{}^{\widehat{\mathcal{o}}}=1+(m_{\psi}c^2)^{\alpha^{(i)}}\sum_{k=0}^6 s_k^{(i)}\gamma^{k-2}+\dots\,.
\end{equation}
\end{subequations}
\begin{table}[t]
\centering
\begin{tabular}{cccccccccccccc}
\toprule
Operator & $i$ & $\alpha^{(i)}$ & $s_0^{(i)}$ & $s_1^{(i)}$ & $s_2^{(i)}$ & $s_3^{(i)}$ & $s_4^{(i)}$ & $s_5^{(i)}$ & $s_6^{(i)}$ \\
\colrule
$\widehat{a}^{(5)\mu}_{\mathrm{eff}}$   & $a^{(5)\mu}$ & 1 & 0                                 & 0 & 0                                                      & $-2(\ring{a}{}^{(5)}_0+\ring{a}{}^{(5)}_2)$ & 0 & 0 & 0 \\
$\widehat{c}^{\,(6)\mu}_{\mathrm{eff}}$ & $c^{(6)\mu}$ & 2 & $-\ring{c}{}^{(6)}_4$ & 0 & $-(\ring{c}{}^{(6)}_2+2\ring{c}{}^{(6)}_4)$ & 0 & $3(\ring{c}{}^{(6)}_0+\ring{c}{}^{(6)}_2+\ring{c}{}^{(6)}_4)$ & 0 & 0 \\
$\widehat{e}^{\,(6)}$    & $e^{(6)}$ & 2 & 0 & 0 & 0 & $2(\ring{e}{}^{(6)}_0+\ring{e}{}^{(6)}_2)$ & 0 & 0 & 0 \\
$\widehat{f}^{(6)}$      & $f^{(6)}$ & 4 & 0 & 0 & $-(1/2)(\ring{f}{}^{(6)}_2)^2$ & 0 & $3\ring{f}{}^{(6)}_2(\ring{f}{}^{(6)}_0+\ring{f}{}^{(6)}_2)$ & 0 & $-(5/2)(\ring{f}{}^{(6)}_0+\ring{f}{}^{(6)}_2)^2$ \\
\botrule
\end{tabular}
\caption{Parameters that are comprised in the modified momentum of \eqref{eq:explicit-result-momentum} and in the first-order
expansion of the quantity $Y$ in \eqref{eq:quantity-y}.}
\label{tab:parameters-momentum}
\end{table}
For {\em CPT}- and Lorentz symmetry conserved, it holds that $Y=1$, i.e., deviations from $Y=1$ would indicate nonvanishing coefficients of
the SME. For the isotropic subsets of the nonminimal operators considered, $Y$ was computed exactly in $\beta\equiv v/c$ and at leading order
in Lorentz violation. Thereby the same parameters $s_k^{(i)}$ that were comprised in the particle momentum of \eqref{eq:explicit-result-momentum}
reappear. In \cite{Staub:1963} measurements for three different values of $\beta$ were performed and we choose $\beta=0.99$.
Assuming that a possible violation of Lorentz invariance hides in the (average) experimental error $\Delta=3.6\times 10^{-4}$ of
\cite{Staub:1963}, we would obtain the sensitivity (at $2\sigma$ level) on the isotropic SME coefficients from the condition
\begin{equation}
\label{eq:equation-for-bounds-1}
\bigg|(m_{\psi}c^2)^{\alpha^{(i)}}\sum_{k=0}^6 s_k^{(i)}\gamma^{k-2}\bigg|<2\Delta\,.
\end{equation}
The expanded inequality can be solved analytically with respect to the SME coefficients where the exact inequality $Y<2\Delta$ is solved numerically
as a crosscheck. The deviation from $Y=1$ lies in the $2\sigma$ interval as long as the coefficients satisfy the constraints given in
\tabref{tab:experimental-bounds-1}.
\begin{table}[t]
\begin{tabular}{cccrclc}
\toprule
Dimension & Sector & Unit & Lower bound & Coefficients & Upper bound & $\overline{E}^{(i)}$ [$E_{\mathrm{Pl}}$] \\
\colrule
$d=5$ & Electron & $\mathrm{GeV}^{-1}$ & $-9.4\times 10^{-2}<$   & $\ring{a}{}^{(5)}_0+\ring{a}{}^{(5)}_2$ & $<1.1\times 10^{-1}$ \\
 & & $E_{\mathrm{Pl}}^{-1}$ & $-1.2\times 10^{18}<$ & $\ring{a}{}^{(5)}_0+\ring{a}{}^{(5)}_2$ & $<1.3\times 10^{18}$ & $8.2\times 10^{-19}$ \\
$d=6$ & Electron & $\mathrm{GeV}^{-2}$ & $-17<$                & $\ring{c}{}^{(6)}_0+\ring{c}{}^{(6)}_2+\ring{c}{}^{(6)}_4$ & $<19$ \\
 & & $E_{\mathrm{Pl}}^{-2}$ & $-2.6\times 10^{39}<$ & $\ring{c}{}^{(6)}_0+\ring{c}{}^{(6)}_2+\ring{c}{}^{(6)}_4$ & $<2.9\times 10^{39}$ & $1.9\times 10^{-20}$ \\
$d=6$ & Electron & $\mathrm{GeV}^{-2}$ & $-1.8\times 10^{2\phantom{0}}<$    & $\ring{e}{}^{(6)}_0+\ring{e}{}^{(6)}_2$ & $<2.1\times 10^2$ \\
 & & $E_{\mathrm{Pl}}^{-2}$ & $-2.8\times 10^{40}<$ & $\ring{e}{}^{(6)}_0+\ring{e}{}^{(6)}_2$ & $<3.1\times 10^{40}$ & $5.9\times 10^{-21}$ \\
$d=6$ & Electron & $\mathrm{GeV}^{-2}$ & $-1.3\times 10^{3\phantom{0}}<$ & $\ring{f}{}^{(6)}_0+\ring{f}{}^{(6)}_2$ & $<1.3\times 10^3$ \\
 & & $E_{\mathrm{Pl}}^{-2}$ & $-1.9\times 10^{41}<$ & $\ring{f}{}^{(6)}_0+\ring{f}{}^{(6)}_2$ & $<1.9\times 10^{41}$ & $2.3\times 10^{-21}$ \\
\botrule
\end{tabular}
\caption{Two-sided bounds for the isotropic parts of the operators $\widehat{a}^{(5)\mu}_{\mathrm{eff}}$, $\widehat{c}^{\,(6)\mu}_{\mathrm{eff}}$,
$\widehat{e}^{\,(6)}$, and $\widehat{f}^{(6)}$ for electrons, obtained from the test of Special-Relativity kinematics in \cite{Staub:1963}.}
\label{tab:experimental-bounds-1}
\end{table}

For the operators $\widehat{m}^{(5)}$, $\widehat{a}^{(5)\mu}$, $\widehat{c}^{\,(6)\mu}$, and $\widehat{e}^{\,(6)}$ the two-sided bounds
are asymmetric since \eqref{eq:equation-for-bounds-1} has two different solutions for positive and negative component coefficients. For
$\widehat{f}^{(6)}$ the bound is symmetric as the corresponding Lorentz-violating contributions are quadratic in the component coefficients.
In \tabref{tab:experimental-bounds-1} the bounds have been expressed via the Planck energy $E_{\mathrm{Pl}}$ as well. Due to their different
mass dimensions, the quality of the constraints is difficult to be compared to each other. For this reason we define the characteristic energy
scale associated with each pair of bounds as
\begin{equation}
\overline{E}^{(i)}\equiv \left[\frac{1}{2}(U^{(i)}-L^{(i)})\right]^{-1/\alpha^{(i)}}\,,
\end{equation}
where $U^{(i)}$ stands for the upper bound, $L^{(i)}$ for the lower bound, and the $\alpha^{(i)}$ are given in \tabref{tab:parameters-momentum}.
The computed values can be found in the last column of \tabref{tab:experimental-bounds-1}. Note that the typical energy scale of the experiment
is given by $\overline{E}^{\mathrm{exp}}\equiv \gamma m_{\psi}c^2\approx 3.0\times 10^{-22}E_{\mathrm{Pl}}$. It is obvious that the bounds lie many orders of
magnitude below the Planck scale. Comparing the characteristic energy scale to the typical energy of the experiment, the bounds are still
better than what would be expected directly from $\overline{E}^{\mathrm{exp}}$. The reason can be found in the high powers of the
Lorentz factor that are associated with the nonminimal coefficients, which increases sensitivity.
Note also that bounds on single coefficients can be obtained directly from the results presented assuming that only one coefficient is nonzero
at a time, cf.~\cite{Kostelecky:2008ts}.

Another possibility of obtaining sensitivities are time-of-flight measurements.
In \cite{Brown:1973xm} an experiment at SLAC is described accelerating electrons until they hit a target to produce bremsstrahlung resulting
in hard gamma rays. Behind the target, the electron beam is deflected in a magnet generating visible synchrotron radiation, in addition. Both
radiation pulses travel long distances before arriving at the detection section. The visible light is detected directly. The hard gamma
rays create positrons in another target where the experiment is arranged such that the latter generate Cherenkov radiation to be detected.
The time differences between the synchrotron pulses and the Cherenkov pulses are measured to test for possible dispersion effects. Furthermore,
in other runs, electrons are not deflected by the first magnet but they travel in a long, straight section before being deflected directly
in the vicinity of the detection section. The emitted synchrotron radiation is detected subsequently. This also allows to compare the
time-of-flight of the electrons and the hard gamma rays.

Few years later, a similar experiment \cite{Guiragossian:1974wp} was performed at SLAC. Again, accelerated electrons hit a target to produce
bremsstrahlung. After doing so, the electrons travel along a straight section where in some runs they are accelerated further. Behind the
straight section both the electrons and the bremsstrahlung gamma rays strike another target to create positrons. Finally, two images are
detected: the first originates from the positrons generated by the electrons and the second from the positrons produced by the gamma pulses.
A crucial device before the detector is an rf separator giving a transverse momentum component to the positrons that depends on their
arrival time. Hence, a different arrival time becomes manifest in a spatial separation of the images detected. The time-of-flight difference
is then obtained from the data collected.

\begin{table}[t]
\centering
\begin{tabular}{ccccccc}
\toprule
$\widehat{\mathcal{o}}$ & $i$ & $\alpha^{(i)}$ & $t_{-1}^{(i)}$ & $t_0^{(i)}$ & $t_1^{(i)}$ & $t_2^{(i)}$ \\
    & &                & $t_3^{(i)}$    & $t_4^{(i)}$ & $t_5^{(i)}$ & $t_6^{(i)}$ \\
\colrule
$\widehat{a}^{(5)\mu}_{\mathrm{eff}}$ & $a^{(5)\mu}$ & 1 & $-\ring{a}{}^{(5)}_2$ & 0 & $3(\ring{a}{}^{(5)}_0+\ring{a}{}^{(5)}_2)$ & 0 \\
                         &  & & $-2(\ring{a}{}^{(5)}_0+\ring{a}{}^{(5)}_2)$ & 0 & 0 & 0 \\
$\widehat{c}^{\,(6)\mu}_{\mathrm{eff}}$ & $c^{(6)\mu}$ & 2 & 0 & $2\ring{c}{}^{(6)}_2+3\ring{c}{}^{(6)}_4$ & 0 & $-(4\ring{c}{}^{(6)}_0+5\ring{c}{}^{(6)}_2+6\ring{c}{}^{(6)}_4)$ \\
                         &   & & 0 & $3(\ring{c}{}^{(6)}_0+\ring{c}{}^{(6)}_2+\ring{c}{}^{(6)}_4)$ & 0 & 0 \\
$\widehat{e}^{\,(6)}$ & $e^{\,(6)}$ & 2 & $\ring{e}{}^{(6)}_2$ & 0 & $-3(\ring{e}{}^{(6)}_0+\ring{e}{}^{(6)}_2)$ & 0 \\
                        & &   & $2(\ring{e}{}^{(6)}_0+\ring{e}{}^{(6)}_2)$ & 0 & 0 & 0 \\
$\widehat{f}^{(6)}$ & $f^{(6)}$ & 4 & 0 & $(\ring{f}{}^{(6)}_2)^2$ & 0 & $-(1/2)\ring{f}{}^{(6)}_2(8\ring{f}{}^{(6)}_0+9\ring{f}{}^{(6)}_2)$ \\
                         & &   & 0 & $3(\ring{f}{}^{(6)}_0+\ring{f}{}^{(6)}_2)(\ring{f}{}^{(6)}_0+2\ring{f}{}^{(6)}_2)$ & 0 & $-(5/2)(\ring{f}{}^{(6)}_0+\ring{f}{}^{(6)}_2)^2$ \\
\botrule
\end{tabular}
\caption{Parameters that are comprised in the modified energy of \eqref{eq:energy-from-lagrangian}.}
\label{tab:parameters-energy}
\end{table}
To use the results of such experiments, we need a relationship between the particle energy and its velocity. Based on the Lagrangians in
\secref{sec:isotropic-parts}, we compute the energy of a particle in terms of its four-velocity components via
\begin{equation}
\label{eq:energy-from-lagrangian}
\ring{E}{}^{\widehat{\mathcal{o}}}=-\frac{\partial \ring{L}{}^{\widehat{\mathcal{o}}}}{\partial u^0}\bigg|_{\substack{u^0=c \\ \mathbf{u}=\mathbf{v}}}=\gamma m_{\psi}c^2\bigg(1+(m_{\psi}c^2)^{\alpha^{(i)}}\sum_{k=-1}^6 t_k^{(i)}\gamma^k\bigg)\,.
\end{equation}
The parameters $t_k^{(i)}$ are stated in \tabref{tab:parameters-energy}. Those parameters in \tabref{tab:parameters-momentum} and
\tabref{tab:parameters-energy} connected to the highest power of the Lorentz factor are equal. This is not surprising since for large $\gamma$
these terms dominate and the particle energy corresponds to the particle momentum to a good approximation.

Equation (\ref{eq:energy-from-lagrangian})
is understood as a relationship that can be solved for $v$ to obtain the velocity as a function of the particle energy. Assuming that a
Lorentz-violating signal hides in the uncertainty of the measurement we can obtain sensitivities on the isotropic, nonminimal coefficients
considered. Thereby we use Table I of \cite{Guiragossian:1974wp}. The first two rows give data of test runs that were performed to check
the potential precision of the experiment, amongst other issues. We choose the data in the third row because for this run, the electrons
have a well-defined energy value of $\unit[15]{GeV}$ and the uncertainty in $\Delta v/c$ is smaller compared to the other data sets. The
corresponding result is $\Delta v/c=(-1.22\pm 1.05)\times 10^{-7}$
where the $2\sigma$ interval reads $\{-3.32,0.88\}\times 10^{-7}$. A Lorentz-violating signal at $2\sigma$ level would have shown up as a
$\Delta v/c$ lying outside of the interval given. This shall be sufficient to obtain further conservative sensitivities on the isotropic
nonminimal coefficients considered in this article.

Therefore, one would have to solve \eqref{eq:energy-from-lagrangian} with respect to $v$ for a particular Lagrangian. However, the polynomials
in $v$ on the right-hand side are of third degree at least, which makes it challenging to solve the equation analytically. A Taylor expansion
in $v/c$ would be unreasonable since the velocities considered lie in the vicinity of the speed of light. This is why the
equation is solved numerically. Thereby, the Lorentz-violating coefficients are scanned through looking for a $\Delta v/c$ that lies outside
of the interval above. For this procedure, we consider only one nonvanishing coefficient during each scan and set the remaining ones to zero.
For each operator, the bounds obtained for $\ring{i}_0$, $\ring{i}_2$ (and $\ring{i}_4$ for $\widehat{c}^{\,(6)\mu}$) are approximately the
same. The reason is that for the bounds the combination of coefficients related to the highest power of the Lorentz factor plays the most
important role. From \tabref{tab:parameters-momentum} it can be deduced that the latter are mere sums of coefficients with a global prefactor.
Therefore, the constraints are obtained for these combinations of coefficients. The results are presented in \tabref{tab:experimental-bounds-2}.

A few remarks on these constraints are in order. First, there are only one-sided bounds in contrast to the previous results of
\tabref{tab:experimental-bounds-1}. The reason is that the deviation of $Y$ in \eqref{eq:quantity-y} from 1 can be both positive and
negative dependent on the sign of the SME coefficients. However, the perturbative form of the Lagrangians given by \eqref{eq:kinematic-lagrangians}
allows subluminal solutions only. The occurrence of the Lorentz factor prohibits superluminal propagation. Besides, this is the reason why
a bound on component coefficients for $\widehat{f}^{(6)}$ cannot be obtained in this context. Second, the current bounds are better than the constraints
of \tabref{tab:experimental-bounds-1} by several orders of magnitude. This has two reasons: the typical energy scale of the experiment,
$\overline{E}^{\mathrm{exp}}\approx 1.2\times 10^{-18}E_{\mathrm{Pl}}$, is larger and there are higher powers of Lorentz factors involved.
The latter especially has a high impact on the constraint for $\widehat{a}^{\,(5)\mu}$.
\begin{table}[t]
\begin{tabular}{cccrclr}
\toprule
Dimension & Sector & Unit & Lower bound & Coefficients & Upper bound & $\overline{E}^{(i)}$ [$E_{\mathrm{Pl}}$] \\
\colrule
$d=5$ & Electron & $\mathrm{GeV}^{-1}$ & $8.0\times 10^{-7}<$   & $\ring{a}{}^{(5)}_0+\ring{a}{}^{(5)}_2$ & \\
 & & $E_{\mathrm{Pl}}^{-1}$ & $9.8\times 10^{12}<$ & $\ring{a}{}^{(5)}_0+\ring{a}{}^{(5)}_2$ & & $1.0\times 10^{-13}$ \\
$d=6$ & Electron & $\mathrm{GeV}^{-2}$ & & $\ring{c}{}^{(6)}_0+\ring{c}{}^{(6)}_2+\ring{c}{}^{(6)}_4$ & $<4.5\times 10^{-7}$ & \\
 & & $E_{\mathrm{Pl}}^{-2}$ & & $\ring{c}{}^{(6)}_0+\ring{c}{}^{(6)}_2+\ring{c}{}^{(6)}_4$ & $<6.7\times 10^{31}$ & $1.2\times 10^{-16}$ \\
$d=6$ & Electron & $\mathrm{GeV}^{-2}$ & & $\ring{e}{}^{(6)}_0+\ring{e}{}^{(6)}_2$ & $<1.6\times 10^{-3}$ & \\
 & & $E_{\mathrm{Pl}}^{-2}$ & & $\ring{e}{}^{(6)}_0+\ring{e}{}^{(6)}_2$ & $<2.4\times 10^{35}$ & $2.0\times 10^{-18}$ \\
\botrule
\end{tabular}
\caption{One-sided bounds for the isotropic parts of the operators $\widehat{a}^{(5)\mu}_{\mathrm{eff}}$, $\widehat{c}^{\,(6)\mu}_{\mathrm{eff}}$,
and $\widehat{e}^{\,(6)}$ for electrons, obtained from the time-of-flight experiment in \cite{Guiragossian:1974wp}. The first is understood as a
lower bound and the last two as upper ones.}
\label{tab:experimental-bounds-2}
\end{table}

\subsection{Comparison to existing constraints}

Only few current constraints exist on nonminimal operators in the fermion sector of the SME. The most important papers that give tabulated
results are the data tables \cite{Kostelecky:2008ts} and the basic article \cite{Kostelecky:2013rta} investigating the nonminimal fermion
sector. The latter only lists astrophysical constraints. Latest bounds were obtained in the muon sector and from spectroscopy of (anti)hydrogen
and other systems, cf.~\cite{Gomes:2014kaa,Kostelecky:2015nma}. All these constraints beat the ones obtained here by many orders of
magnitude. On the one hand, this shows the potential of astrophysical experiments involving large propagation distances and/or high energies.
On the other hand, it demonstrates the high precision of spectroscopy experiments, which mainly rely on comparing frequencies. So both
classes of experiments outreach purely kinematic tests of Lorentz and {\em CPT}-invariance. Nevertheless, it is reasonable to have a
quantitative comparison of the different experimental techniques available.

\section{Finsler structures and their properties}
\label{sec:finsler-structures-properties}

After dealing with mainly phenomenological issues in the past sections we are now interested in a couple of theoretical questions.
One of the essential results of \cite{Kostelecky:2010hs,Kostelecky:2011qz} is that classical Lagrangians associated to the SME fermion
sector can be promoted to Finsler structures. Considering a manifold $M$, a Finsler structure can be understood as the generalization of
the integrand in the path length functional. In general, the latter depends both on the coordinate $x\in M$ and a direction $\mathbf{y}\in T_xM$
where $T_xM$ is the tangent space of the manifold at $x$. Denoting the tangent bundle with $TM\equiv \cup_{x\in M}T_xM$, a Finsler structure
$F=F(x,\mathbf{y}):TM\mapsto [0,\infty)$ is characterized by a set of properties, cf.~\cite{Antonelli:1993,Bao:2000}:

\begin{itemize}

\item[1)] $F(x,\mathbf{y})>0$ for $\mathbf{y}\in TM\setminus\{0\}$,
\item[2)] $F(x,\mathbf{y})\in C^{\infty}$ for $\mathbf{y}\in TM\setminus \{0\}$,
\item[3)] positive homogeneity in $y$, i.e., $F(x,\lambda \mathbf{y})=\lambda F(x,\mathbf{y})$ for $\lambda>0$, and
\item[4)] the derived metric (Finsler metric)
\begin{equation}
\label{eq:derived-metric}
g_{ij}\equiv \frac{1}{2}\frac{\partial^2 F^2}{\partial y^i\partial y^j}\,,
\end{equation}
is positive definite for $\mathbf{y}\in TM\setminus\{0\}$.

\end{itemize}
To find a connection between a classical Lagrangian and such a Finsler structure two procedures have been proposed \cite{Kostelecky:2011qz}.
The first is to set $u^0=0$, i.e., to restrict the Lagrangian to its spatial subset. The second procedure bears resemblance to a Wick rotation.
It requires to introduce a four-dimensional vector $\mathbf{y}=(y^1,y^2,y^3,y^4)^T$ that is linked to the
velocity four-vector via $u^0=\mathrm{i}y^4$, $u^i=y^i$ for $i\in \{1\dots 3\}$. The Lorentz-violating coefficients have to be
treated in a similar manner in this context. We will follow the second path that will be slightly adapted to serve its purpose in
the nonminimal fermion sector of the SME.

\subsection{Finsler structure for \texorpdfstring{$\boldsymbol{\widehat{m}^{(5)}}$}{m5-hat}}
\label{sec:derivation-finsler-structure-m5}

Let us first consider $\widehat{m}^{(5)}$ with
the result given in \eqref{eq:lagrangian-m5-complete}. A Finsler structure is expected to be found by the above replacement rules for
the four-velocity with further rules applied to the Lorentz-violating controlling coefficients. For this case, each coefficient is multiplied
by a factor of $(-\mathrm{i})^q$ where $q$ is the number of indices equal to zero. This means that a coefficient gets one factor of
$-\mathrm{i}$ per timelike index and each of the latter is replaced by 4:
\begin{subequations}
\begin{align}
m^{(5)}_{ij}&\mapsto m^{(5)}_{ij}\,,\quad m^{(5)}_{0i} \mapsto -\mathrm{i}m^{(5)}_{4i}\,,\quad m^{(5)}_{i0}\mapsto -\mathrm{i}m^{(5)}_{i4}\,, \\[2ex]
m^{(5)}_{00}&\mapsto (-\mathrm{i})^2m^{(5)}_{44}=-m^{(5)}_{44}\,.
\end{align}
\end{subequations}
In principle, this corresponds to introducing controlling coefficients in Euclidean space. Partially, the latter are not equivalent
to the original Lorentz-violating coefficients anymore. After all, the concept of Lorentz symmetry is lost in Euclidean space.
Nevertheless, for simplicity the new coefficients will still be referred to as the Lorentz-violating component coefficients.
Now, the Finsler structure to-be has the following form:
\begin{equation}
\label{eq:finsler-structure-m5}
F^{\widehat{m}^{(5)}}\equiv -\frac{\mathrm{i}}{m_{\psi}}L^{\widehat{m}^{(5)}+}\big|_{\mathrm{Wick}}=\sqrt{\mathbf{y}^2}-\frac{m_{\psi}}{\sqrt{\mathbf{y}^2}}\widehat{M}^{(5)}_{\ast}\,,\quad \widehat{M}^{(5)}_{\ast}\equiv \sum_{i,j=1}^4 m^{(5)}_{ij}y^iy^j\,,
\end{equation}
where by the subscript ``Wick'' we indicate that the replacement rules for $u^{\mu}$ and the Lorentz-violating coefficients have to be applied.
Note that this result is defined in Euclidean space, where $F^{\widehat{m}^{(5)}}$ does not depend on the position~$x$. Now the immediate question is
whether the latter fulfills the properties of a Finsler structure. Here it must be kept in mind that all Lagrangians obtained in
the current paper are perturbative results, i.e., they are only valid for a sufficiently small dimensionless product of powers of
the particle mass $m_{\psi}$ and controlling coefficients. Hence, the latter $F^{\widehat{m}^{(5)}}$ must be understood as a perturbative result as well. As
long as the second term is much smaller than the first, property (1) is granted. This even holds for large components of $y$. For
example, let us suppose that there is one component $y^a$ that dominates all the others. Then $F^{\widehat{m}^{(5)}}$ has the following asymptotic behavior
(equal indices are not summed over):
\begin{equation}
F^{\widehat{m}^{(5)}}(y^a)\approx \sqrt{(y^a)^2}-\frac{m_{\psi}}{\sqrt{(y^a)^2}}m^{(5)}_{aa}(y^a)^2=\sqrt{(y^a)^2}\left(1-m_{\psi}m^{(5)}_{aa}\right)\,,
\end{equation}
which is still positive for $m_{\psi}m^{(5)}_{aa}\ll 1$. An analogous argument holds for several dominant coefficients. Concerning (2),
the function $F(y)$ involves the square root, its inverse, and polynomials in $y^i$. Thus, it is $C^{\infty}\setminus \{0\}$.
Positive homogeneity, i.e., (3) can be shown by direct computation:
\begin{equation}
F^{\widehat{m}^{(5)}}(\lambda\mathbf{y})=\sqrt{\lambda^2\mathbf{y}^2}-\frac{m_{\psi}}{\sqrt{\lambda^2\mathbf{y}^2}}\sum_{i,j=1}^4 m^{(5)}_{ij}(\lambda y^i)(\lambda y^j)=\lambda F^{\widehat{m}^{(5)}}(\mathbf{y})\,,
\end{equation}
for $\lambda>0$. By the way, this property ensures the nonstandard contribution in $F(\mathbf{y})$ to be perturbative even for large
components $y^i$. To show property (4) one has to compute the derived metric $g_{ij}$ of \eqref{eq:derived-metric}, which will be
postponed to a later part of the paper.

\subsection{Finsler structure for \texorpdfstring{$\boldsymbol{\widehat{a}^{(5)\mu}}$}{a5mu-hat}}

As a next step, we would like to assign a Finsler structure to the Lagrangian of the operator $\widehat{a}^{(5)}$ given by
\eqref{eq:lagrangian-a5}. The procedure employed for $\widehat{m}^{(5)}$ does not work here since the Lorentz-violating contribution
involves a trilinear combination of components of $\mathbf{y}$. Therefore, it has to be adapted where suitable replacement rules are
given as follows:
\begin{subequations}
\begin{align}
a^{(5)}_{ijk}&\mapsto -\mathrm{i}a^{(5)}_{ijk}\,,\quad a^{(5)}_{0ij}\mapsto -a^{(5)}_{4ij}\,,\quad a^{(5)}_{i0j}\mapsto -a^{(5)}_{i4j}\,,\quad a^{(5)}_{ij0}\mapsto -a^{(5)}_{ij4}\,, \\[2ex]
a^{(5)}_{00i}&\mapsto \mathrm{i}a^{(5)}_{44i}\,,\quad a^{(5)}_{0i0}\mapsto \mathrm{i}a^{(5)}_{4i4}\,,\quad a^{(5)}_{i00}\mapsto \mathrm{i}a^{(5)}_{i44}\,,\quad a^{(5)}_{000}\mapsto a^{(5)}_{444}\,.
\end{align}
\end{subequations}
Hence, in contrast to the case of $\widehat{m}^{(5)}$, there is one factor of $\mathrm{i}$ per each spacelike coefficient. Furthermore,
each timelike coefficient is replaced by 4 just as before. The resulting Finsler structure to-be can then be cast into the form
\begin{equation}
\label{eq:finsler-structure-a5}
F^{\widehat{a}^{(5)\mu}}\equiv -\frac{\mathrm{i}}{m_{\psi}}L^{\widehat{a}^{(5)}+}\big|_{\mathrm{Wick}}=\sqrt{\mathbf{y}^2}-\frac{m_{\psi}}{\mathbf{y}^2}\widehat{A}^{(5)}_{\ast}\,,\quad \widehat{A}^{(5)}_{\ast}\equiv\sum_{i,j,k=1}^4 a^{(5)}_{ijk}y^iy^jy^k\,.
\end{equation}
Properties (1) and (2) are valid in analogy to $\widehat{m}^{(5)}$ as long as Lorentz violation is perturbative. Positive homogeneity
is granted by the functional behavior in $\mathbf{y}$:
\begin{equation}
F^{\widehat{a}^{(5)\mu}}(\lambda\mathbf{y})=\sqrt{\lambda^2\mathbf{y}^2}-\frac{m_{\psi}}{\lambda^2\mathbf{y}^2}\sum_{i,j,k=1}^4 a^{(5)}_{ijk}(\lambda y^i)(\lambda y^j)(\lambda y^k)=\lambda F^{\widehat{a}^{(5)\mu}}(\mathbf{y})\,,
\end{equation}
for $\lambda>0$. The corresponding derived metric $g_{ij}$ will be obtained later.

\subsection{Generic Finsler structure}
\label{sec:generic-finsler-structure}

A Finsler structure to-be can be derived for $\widehat{c}^{\,(6)\mu}$ according to the method for $\widehat{m}^{(5)}$ used in
\secref{sec:derivation-finsler-structure-m5} as the component coefficients of $\widehat{c}^{\,(6)\mu}$ have an even number of indices.
For both $\widehat{e}^{\,(6)}$ and $\widehat{f}^{(6)}$ the technique for $\widehat{a}^{(5)\mu}$ works well since the corresponding
component coefficients have an odd number of indices. The resulting $F(\mathbf{y})$ have the general form
\begin{equation}
\label{eq:generic-finsler-structure}
F^{\widehat{\mathcal{o}}}=\sqrt{\mathbf{y}^2}-m_{\psi}^{l_1^{(i)}}\frac{l_2^{(i)}}{(\mathbf{y}^2)^{l_3^{(i)}}}\widehat{O}_{\ast}^{(i)}\,,
\end{equation}
where $(i)$ stands for $m^{(5)}$, $a^{(5)\mu}$, etc., i.e., effective coefficients will not be used anymore. The values of the
parameters $l_1^{(i)}\dots l_3^{(i)}$ can be found in
\tabref{tab:numbers-finsler-structures}. For each operator, the generic function $\widehat{O}_{\ast}^{(i)}$
is a proper contraction of the controlling coefficients (with each timelike index replaced by 4) and components of $\mathbf{y}$
based on the Euclidean metric (cf.~\eqref{eq:finsler-structure-m5} for $\widehat{m}^{(5)}$ and \eqref{eq:finsler-structure-a5} for
$\widehat{a}^{(5)\mu}$). Care has to be taken for $\widehat{f}^{\,(6)}$ whose contraction must be squared in addition.
\begin{table}[b]
\centering
\begin{tabular}{cccccc}
\toprule
Operator & $i$                      & $l_1^{(i)}$ & $l_2^{(i)}$ & $l_3^{(i)}$ \\
\colrule
$\widehat{m}^{(5)}$      & $m^{(5)}$ & 1 & 1      & 1/2 \\
$\widehat{a}^{(5)\mu}$   & $a^{(5)\mu}$ & 1 & 1      & 1   \\
$\widehat{c}^{\,(6)\mu}$ & $c^{\,(6)\mu}$ & 2 & 1      & 3/2 \\
$\widehat{e}^{\,(6)}$    & $e^{\,(6)}$ & 2 & $-1$   & 1   \\
$\widehat{f}^{(6)}$      & $f^{(6)}$ & 4 & $-1/2$ & 5/2 \\
\botrule
\end{tabular}
\caption{Parameters for the generic Finsler structure of \eqref{eq:generic-finsler-structure}.}
\label{tab:numbers-finsler-structures}
\end{table}

Now this general form of $F^{(i)}$ is taken to compute the derived metric. For brevity, we drop the indices of $l_1^{(i)}$ and
$\widehat{O}_{\ast}^{(i)}$, and we introduce the function $f\equiv l_2^{(i)}/(\mathbf{y}^2)^{l_3^{(i)}}$. Based on its definition
in \eqref{eq:derived-metric}, the derived metric takes the form
\begin{subequations}
\label{eq:derived-metric-generic}
\begin{align}
g_{ij}&=\delta_{ij}-\frac{m_{\psi}^{l_1}}{\sqrt{y^2}}\mathcal{G}_{ij}^{(1)}+m_{\psi}^{2l_1}\mathcal{G}_{ij}^{(2)}\,, \displaybreak[0]\\[2ex]
\label{eq:derived-metric-generic-function-g1}
\mathcal{G}_{ij}^{(1)}&=\widehat{O}_{\ast}\bigg[f\bigg(\delta_{ij}-\frac{y_iy_j}{y^2}\bigg)+\bigg(y_i\frac{\partial f}{\partial y^j}+(i\leftrightarrow j)\bigg)+y^2\frac{\partial^2f}{\partial y^i\partial y^j}\bigg] \notag \\
&\phantom{{}={}}+\bigg\{\frac{\partial\widehat{O}_{\ast}}{\partial y^i}\bigg(y^jf+y^2\frac{\partial f}{\partial y^j}\bigg)+(i\leftrightarrow j)\bigg\}+y^2f\frac{\partial\widehat{O}_{\ast}}{\partial y^i\partial y^j}\,, \displaybreak[0]\\[2ex]
\mathcal{G}_{ij}^{(2)}&=(\widehat{O}_{\ast})^2\left(\frac{\partial f}{\partial y^i}\frac{\partial f}{\partial y^j}+f\frac{\partial^2f}{\partial y^i\partial y^j}\right)+2f\widehat{O}_{\ast}\bigg(\frac{\partial f}{\partial y^i}\frac{\partial\widehat{O}_{\ast}}{\partial y^j}+(i\leftrightarrow j)\bigg) \notag \\
&\phantom{{}={}}+f^2\bigg(\frac{\partial\widehat{O}_{\ast}}{\partial y^i}\frac{\partial\widehat{O}_{\ast}}{\partial y^j}+\widehat{O}_{\ast}\frac{\partial^2\widehat{O}_{\ast}}{\partial y^i\partial y^j}\bigg)\,.
\end{align}
\end{subequations}
It is reasonable to arrange all terms according to the power of the particle mass they are multiplied with. A couple of remarks on
the result are in order. First, for vanishing Lorentz violation, both $\mathcal{G}^{(1)}$ and $\mathcal{G}^{(2)}$ vanish giving the
standard result $g_{ij}=\delta_{ij}$. Second, by direct inspection, the symmetry of $g_{ij}$ in its indices can be checked.
Third, $\mathcal{G}^{(1)}$ is of first order in Lorentz violation whereas $\mathcal{G}^{(2)}$ is of second order. Based on the
perturbative nature of Lorentz violation, the derived metric is a perturbation of the identity matrix multiplied by the particle
square. The determinants of all main minors of a perturbed identity matrix lie in the vicinity of 1, since all these matrices
have coefficients of the form ``$1+\text{perturbation}$'' on their diagonals. This renders $g_{ij}$ positive definite. Thus, under
the assumptions made, \eqref{eq:generic-finsler-structure} defines a Finsler structure.

Now some of the mathematical properties of this Finsler structure can be investigated. The first is to check whether
\eqref{eq:generic-finsler-structure} is just the usual Riemannian structure that was written in a complicated form. This check
can be accomplished with the Cartan torsion $C_{ijk}$ and the mean Cartan torsion $I_i$ that are defined by \cite{Bao:2000}:
\begin{equation}
\label{eq:cartan-torsion-definition}
C_{ijk}\equiv\frac{F}{4}\frac{\partial^3F^2}{\partial y^i\partial y^j\partial y^k}=\frac{F}{2}\frac{\partial g_{ij}}{\partial y^k}\,,\quad I_i\equiv g^{jk}C_{ijk}\,,\quad (g^{ij})\equiv (g_{ij})^{-1}\,.
\end{equation}
Hence, the Cartan torsion is obtained from the Finsler metric by computing an additional partial derivative with respect to $\mathbf{y}$.
The mean Cartan torsion follows from the Cartan torsion by a suitable contraction with the inverse Finsler metric $g^{ij}$. A
result for the latter will not be given as it is highly complicated and not illuminating. It is most reasonable to obtain the
inverse Finsler metric with a computer algebra system.
According to a theorem by Deicke \cite{Deicke:1953}, a Finsler structure defines a Riemannian space if and only if the Cartan
torsion (or the mean Cartan torsion) vanishes. For the generic Finsler structure of \eqref{eq:generic-finsler-structure} the Cartan
torsion is quite lengthy, which is why it is shown in \appref{sec:cartan-matsumoto-torsion} explicitly. It decomposes into three main
parts. The first two are of first order in Lorentz violation and the third is of second order.

The general result of \eqref{eq:cartan-torsion-generic} can be used to obtain the Cartan torsions for each of the nonminimal cases considered. Interestingly,
for $\widehat{m}^{(5)}$ it was found that the first-order term vanishes, which was surprising. The reason is that for the specific
function $f=1/\sqrt{\mathbf{y}^2}$ the second part of the Finsler metric, \eqref{eq:derived-metric-generic-function-g1}, is
independent of $\mathbf{y}$, which makes its first derivative vanish. This is a property of the particular $\widehat{m}^{(5)}$
and it does not hold for the other operators. It means that the Finsler structure that is linked to $\widehat{m}^{(5)}$ in the
SME fermion sector deviates from Riemannian geometry only at second order in the controlling coefficients. To understand whether
and what implications this has for physics would be a worthwhile problem to study.

The Finsler structures of the remaining operators have nonvanishing Cartan torsions at first order in Lorentz violation except of
$\widehat{f}^{\,(6)}$, which is clear as the leading-order term of the latter is quadratic at the first place. Hence, none of these
Finsler spaces is Riemannian. Another interesting quantity useful for classification purposes is the Matsumoto torsion:
\begin{equation}
\label{eq:matsumoto-torsion-definition}
M_{ijk}\equiv C_{ijk}-\frac{1}{n+1}(I_ih_{jk}+I_jh_{ik}+I_kh_{ij})\,,\quad h_{ij}\equiv F\frac{\partial^2F}{\partial y^i\partial y^j}\,.
\end{equation}
The latter is comprised of the Cartan torsion, the mean Cartan torsion, and the angular metric tensor $h_{ij}$ where $n$ is the dimension of
the Finsler space considered. The full result will not be given as it is very lengthy. However, it can be constructed from $C_{ijk}$
in \eqref{eq:cartan-torsion-generic} and $h_{ij}$ given by \eqref{eq:average-curvature-generic}. According to a theorem by
Matsumoto and H\={o}j\={o} \cite{Matsumoto:1978}, a Finsler structure is of Randers or Kropina-type\footnote{
A Kropina structure is defined by $F(x,y)=\alpha^2/\beta$ with $\alpha=\sqrt{a_{ij}(x)y^iy^j}$ and $\beta=b_i(x)y^i$ where $a_{ij}(x)$ is a
Riemannian metric and $b_i(x)$ a 1-form.} if and only if the Matsumoto torsion
vanishes. The Matsumoto torsion for $\widehat{m}$ and $\widehat{f}$ each delivers contributions at second order in Lorentz violation
whereas for $\widehat{a}^{(5)\mu}$, $\widehat{c}^{\,(6)\mu}$, and $\widehat{e}^{\,(6)}$ they are of first order. Especially the result
for $\widehat{m}^{(5)}$ is again interesting as it shows that for the associated Finsler structure both deviations from a Riemannian
and a Randers/Kropina structure are of higher order.

\subsection{Covariantly constant background coefficients and Berwald spaces}

This last section shall give us further insight on the mathematical properties of the Finsler structures previously introduced. Therefore,
the generic Finsler structure of \eqref{eq:generic-finsler-structure} is promoted to exist on a curved manifold, i.e., the Euclidean
metric is replaced by a Riemannian metric $r_{ij}=r_{ij}(x)$ and the component coefficients are taken to be position-dependent functions:
\begin{equation}
\label{eq:riemann-finsler-space}
F^{\widehat{\mathcal{o}}}\mapsto F^{\widehat{\mathcal{o}}}(x)=\sqrt{r_{ij}(x)y^iy^j}-m_{\psi}^{l_1^{(i)}}\frac{l_2^{(i)}}{(r_{ij}(x)y^iy^j)^{l_3^{(i)}}}\widehat{O}_{\ast}^{(i)}(x)\,.
\end{equation}
The fermion mass $m_{\psi}$ is kept constant. In principle, it could be absorbed by the component coefficients but we will leave it as is.
In $\widehat{O}_{\ast}^{(i)}(x)$ there is no $r_{ij}$ as long as the indices
of the (originally) Lorentz-violating component coefficients are kept as lower ones. With this Riemann-Finsler structure at hand, we
want to understand an aspect whose importance especially for $b$ space was pointed out in \cite{Kostelecky:2011qz}. Consider coefficients
that are parallel with respect to the Riemannian metric $r_{ij}$, which means that their covariant derivatives based on $r_{ij}$ are
supposed to vanish. In \cite{Kostelecky:2011qz} such spaces were demonstrated to be of Berwald-type, which will be elaborated on after
introducing a set of important quantities in this context.

In what follows, we use the conventions and definitions of \cite{Bao:2000}. First of all, the formal Christoffel symbols of the
second kind for the Riemannian metric tensor $r_{ij}$ and the Finsler metric tensor $g_{ij}$ are obtained as follows:
\begin{subequations}
\begin{align}
\widetilde{\gamma}^i_{\phantom{i}jk}\equiv\frac{1}{2}r^{il}\left(\frac{\partial r_{lk}}{\partial x^j}+\frac{\partial r_{lj}}{\partial x^k}-\frac{\partial r_{jk}}{\partial x^l}\right)\,, \\[2ex]
\gamma^i_{\phantom{i}jk}\equiv\frac{1}{2}g^{il}\left(\frac{\partial g_{lk}}{\partial x^j}+\frac{\partial g_{lj}}{\partial x^k}-\frac{\partial g_{jk}}{\partial x^l}\right)\,.
\end{align}
\end{subequations}
An important quantity for studying the geodesic equations in a Riemann-Finsler space are the spray coefficients
$G^i\equiv\gamma^i_{\phantom{i}jk}y^jy^k$. Also, the Chern connection is needed. It comprises the nonlinear connection
$N^i_{\phantom{i}j}$ and can be obtained according to
\begin{subequations}
\begin{align}
\Gamma^i_{\phantom{i}jk}&\equiv\frac{1}{2}g^{il}\left(\frac{\delta g_{lk}}{\delta x^j}+\frac{\delta g_{lj}}{\delta x^k}-\frac{\delta g_{jk}}{\delta x^l}\right)\,, \\[2ex]
\frac{\delta}{\delta x^k}&\equiv\frac{\partial}{\partial x^k}-N^i_{\phantom{i}k}\frac{\partial}{\partial y^i}\,,\quad N^i_{\phantom{i}j}\equiv \frac{1}{2}\frac{\partial G^i}{\partial y^j}\,.
\end{align}
\end{subequations}
Note the similar structure of the Christoffel symbols and the Chern connection.
Coming back to a Berwald space, the latter is a Finsler space whose Chern connection coefficients $\Gamma^i_{\phantom{i}jk}$
in natural coordinates do not have any dependence on $\mathbf{y}$. Under that condition, the object known as the h-v part of the
Chern curvature vanishes identically:
\begin{equation}
P_{j\phantom{i}kl}^{\phantom{j}i}\equiv -F\frac{\partial\Gamma^i_{\phantom{i}jk}}{\partial y^l}=0\,.
\end{equation}
The second derivatives of the spray coefficients with respect to the components of $\mathbf{y}$ vanish as well, which is
why $\gamma^i_{\phantom{i}jk}=\widetilde{\gamma}^i_{\phantom{i}jk}$. Hence, the Christoffel symbols of the Finsler metric
tensor reduce to the usual Riemannian Christoffel symbols. In turn, this is equivalent to a vanishing Berwald curvature~${}^bP_{j\phantom{i}kl}^{\phantom{j}i}$:
\begin{equation}
{}^bP_{j\phantom{i}kl}^{\phantom{j}i}\equiv-F\frac{\partial {}^b\Gamma^i_{\phantom{i}jk}}{\partial y^l}=0\,,\quad {}^b\Gamma^i_{\phantom{i}jk}\equiv \frac{1}{2}\frac{\partial^2G^i}{\partial y^j\partial y^k}\,,
\end{equation}
with the Berwald connection ${}^b\Gamma^i_{\phantom{i}jk}$. The tangent spaces of a Berwald space are linearly isometric to
a common Minkowski space.\footnote{Here the term ``Minkowski space'' is not to be confused with physical spacetime but with
a mathematical space endowed with a Minkowski norm, cf.~\cite{Bao:2000}.} All these properties mentioned make Berwald spaces very special Finsler spaces ``that are more
properly Finslerian, but only slightly so,'' which is an accurate statement taken from \cite{Bao:2000}.

One of the most important results of \cite{Kostelecky:2011qz} is the proof that $b$ space is Berwald if its component
coefficients are covariantly constant. It was then conjectured that this holds in general and also in the opposite direction,
i.e., any Berwald space would have $r$-parallel coefficients associated to it. For Randers space such a statement is known to
hold in both directions, cf.~the references given in \cite{Kostelecky:2011qz}. Therefore, with the above definitions
at hand we would like to investigate whether \eqref{eq:riemann-finsler-space} also gives rise to a Berwald space in case the
component coefficients are covariantly conserved.
There exist several possibilities of performing the calculation. We decide to check the spray coefficients
for a possible dependence on $\mathbf{y}$. This is reasonable as due to the homogeneity properties of the Finsler
structure, the spray coefficients are supposed to collapse in complexity when the Christoffel symbols are contracted with
two components of $\mathbf{y}$. Hence, the spray coefficients (with their index pulled down) can be brought into the form
\begin{align}
\label{eq:spray-coefficients-index-down}
G_i&=\gamma_{ijk}y^jy^k=\frac{1}{2}\left(\frac{1}{2}\frac{\partial^2F^2}{\partial y^i\partial x^k}y^k+\frac{1}{2}\frac{\partial^2F^2}{\partial y^i\partial x^l}y^l-\frac{\partial F^2}{\partial x^i}\right) \notag \\
&=\frac{1}{2}\left[2\frac{\partial}{\partial x^k}\left(F\frac{\partial F}{\partial y^i}\right)y^k-2F\frac{\partial F}{\partial x^i}\right]=\left(\frac{\partial F}{\partial x^k}\frac{\partial F}{\partial y^i}+F\frac{\partial^2F}{\partial y^i\partial x^k}\right)y^k-F\frac{\partial F}{\partial x^i}\,.
\end{align}
Useful formulas to compute derivatives of \eqref{eq:riemann-finsler-space} are
\begin{subequations}
\begin{align}
\frac{\partial f(\mathbf{y}^2)}{\partial x^k}&=f'(\mathbf{y}^2)\frac{\partial r_{ij}}{\partial x^k}y^iy^j\,, \\[2ex]
\frac{\partial f(\mathbf{y}^2)}{\partial y^k}&=2f'(\mathbf{y}^2)r_{ij}y^j\,.
\end{align}
\end{subequations}
The procedure is to obtain $\gamma_{ijk}y^jy^k$ according to \eqref{eq:spray-coefficients-index-down} and to pull its
free index up with the inverse Finsler metric $g^{ij}$. Finally, the property of component coefficients parallel to $r_{ij}$
has to be employed. This means that their covariant derivatives $D_i$ based on the Riemannian metric $r_{ij}$ vanishes. Explicitly,
for a two-, three-, and four-index object that statement reads as follows:
\begin{subequations}
\begin{align}
D_km^{(5)}_{ij}&=\frac{\partial m^{(5)}_{ij}}{\partial x^k}-(\widetilde{\gamma}^l_{\phantom{l}ki}m^{(5)}_{lj}+\widetilde{\gamma}^l_{\phantom{d}kj}m^{(5)}_{il})\overset{!}{=}0\,, \\[2ex]
D_la^{(5)}_{ijk}&=\frac{\partial a^{(5)}_{ijk}}{\partial x^l}-(\widetilde{\gamma}^{m}_{\phantom{m}li}a^{(5)}_{mjk}+\widetilde{\gamma}^{m}_{\phantom{m}lj}a^{(5)}_{imk}+\widetilde{\gamma}^{m}_{\phantom{m}lk}a^{(5)}_{ijm})\overset{!}{=}0\,, \\[2ex]
D_mc^{(6)}_{ijkl}&=\frac{\partial c^{(6)}_{ijkl}}{\partial x^m}-(\widetilde{\gamma}^{n}_{\phantom{n}mi}c^{(6)}_{njkl}+\widetilde{\gamma}^{n}_{\phantom{n}mj}c^{(6)}_{inkl}+\widetilde{\gamma}^{n}_{\phantom{n}mk}c^{(6)}_{ijn l}+\widetilde{\gamma}^{n}_{\phantom{n}ml}c^{(6)}_{ijkn})\overset{!}{=}0\,,
\end{align}
\end{subequations}
where all coefficients are understood to depend on $x$.
Solving these relations with respect to the partial derivative of the coefficients leads to a set of replacement rules.
Thereby, all partial derivatives of component coefficients that occur in $G^i$
are replaced by the corresponding sums of Christoffel symbols $\widetilde{\gamma}^i_{\phantom{i}jk}$ contracted with
component coefficients. Even for $\widehat{m}^{(5)}$ the computation is arduous and it seems to be challenging to reduce
it to an elegant formula such as Eq.~(25) in \cite{Kostelecky:2011qz}, which is valid for $b$ space. However, note that
the Finsler metric in \eqref{eq:derived-metric-generic} and the Cartan torsion of \eqref{eq:cartan-torsion-generic} are much
more involved than their $b$-space counterparts given by Eqs.~(10), (13) in the latter paper. The largest computational issues
are related to the inverse Finsler metric that cannot be expressed in a compact form. This only allows us to demonstrate
numerically (with 100-digit accuracy) that the spray coefficients based on the Finsler metric are the same as the spray
coefficients based on $r_{ij}$.

Therefore, there is strong evidence that in case the component coefficients are parallel with respect to the Riemannian metric $r_{ij}$, the Finsler structure
of \eqref{eq:riemann-finsler-space} is Berwald. This means that the corresponding geodesic equations are not influenced by
the Lorentz-violating background but they are just governed by the Riemannian metric $r_{ij}$. The latter especially holds in
Euclidean space when the coefficients are constant. The corresponding property in Minkowski spacetime is that a classical
particle still moves on a straight line for a constant Lorentz-violating background.

This result further supports the
conjecture made in \cite{Kostelecky:2011qz}. The conjecture is interesting since it could provide a criterion allowing us to
decide when geodesic motion of a particle in the presence of Lorentz violation differs from the Lorentz-invariant
case. In \cite{Kostelecky:2003fs} it was noted that covariantly conserved coefficients are related to only a very special class
of curved manifolds such as parallelizable ones. However, even if a manifold is not parallelizable globally the result obtained
would still be valid locally. Besides, having a manifold that differs from a parallelizable one only slightly it is reasonable
to assume that Lorentz-violating deviations from conventional particle motion are suppressed.

Covariantly constant coefficients may still show up in particle motion if there exists an additional electric or magnetic field,
as pointed out for Minkowski spacetime in \cite{Schreck:2014hga,Schreck:2014ama} and at the beginning of
\secref{sec:experimental-sensitivity}. In fact, the fields deliver additional degrees of freedom to reveal the $r$-parallel
Lorentz-violating background. The Lorentz-violating effects on particle motion are then suppressed by the field strengths
involved, though.

\section{Conclusions and outlook}
\label{sec:conclusions}

In the current paper, we obtained classical point-particle equivalents to a series of operators of the nonminimal SME fermion
sector. The calculations were carried out at first order in Lorentz violation and they relied on the algebraic concept of Gr\"{o}bner
bases to solve the resulting nonlinear systems of equations. All Lagrangians are characterized by the standard square root term
plus a Lorentz-violating contribution that comprises suitable contractions of the controlling coefficients with four-velocity
components. These first-order expansions are much more illuminating and practical than the nonperturbative result found for the
nonminimal, isotropic coefficient $m^{(5)}_{00}$ in \cite{Schreck:2014hga}. This shows that in the context of classical Lagrangians
of the nonminimal SME, it seems to be more reasonable to perform studies based on perturbative Lorentz violation.

These Lagrangians served as a basis for phenomenological studies. Thereby, we relied on classical, kinematic tests of Special
Relativity carried out in the middle of the past century. The first type of experiments, which we considered, measured the mass
increase of an electron for relativistic velocities. In the second type the time-of-flight for hard gamma rays was compared to
the time-of-flight for relativistic electrons. Both types of experiments confirmed the validity of Special Relativity to a certain
accuracy. This allowed for deriving constraints on the isotropic component coefficients of the nonminimal operators considered.
These bounds are still far away from the Planck scale, which leaves plenty of room for improvement by future kinematic experiments.

Finally, it was demonstrated that the classical Lagrangians are connected to Finsler structures in analogy to the minimal SME
fermion sector. Expressions for the Finsler metric, the Cartan torsion and the Matsumoto torsion were obtained, which allowed
for their classification. The structures considered are neither Riemannian nor of Randers or Kropina-type. An interesting and unexpected
result is that the structure associated to the dimension-5 operator $\widehat{m}$ deviates from a Riemannian or a Randers/Kropina
structure only at second order in Lorentz violation. This result may have important implications for the motion of classical
particles in a gravitational background in case they are subject to such a particular type of Lorentz violation. Last but not
least, it was found that the Finsler structures promoted to a curved background are of Berwald-type in case the component
coefficients are covariantly constant with respect to the Riemannian metric of the curved manifold. This provides a link between
the nature of coefficients and the geometry of the Finsler space associated. It also supports the general conjecture made in
\cite{Kostelecky:2011qz} on covariantly conserved coefficients. Additionally, another interesting thought was brought up in
\cite{Kostelecky:2011qz}. For $r$-parallel coefficients there could be a coordinate redefinition that allows to remove these
coefficients from the Finsler structure. This would be reminiscent of field redefinitions eliminating unphysical coefficients from
the SME.

There are a couple of interesting problems still to be solved. The first question is whether and how Lagrangians
for the nonminimal operators $\widehat{b}^{(5)\mu}$, $\widehat{d}^{\,(6)\mu}$, $\widehat{g}^{(6)\mu\nu}$, and $\widehat{H}^{(5)\mu\nu}$
can be derived. These cases are calculationally more involved than the operators considered within the paper. Second, the
phenomenological studies performed relied on experiments testing the kinematic laws of particles directly. Applying the same
technique to alternative experiments may provide fruitful opportunities of obtaining an even better set of constraints. The
third issue is to study particle propagation in a curved background based on the classical Lagrangians calculated.

\section{Acknowledgments}

It is a pleasure to thank V.~A.~Kosteleck\'{y} for very useful discussions that helped to improve the quality of the article.
This work was partially funded by the Brazilian foundation FAPEMA.

\newpage
\begin{appendix}
\numberwithin{equation}{section}

\section{Obtaining the classical Lagrangian of $\boldsymbol{\widehat{m}^{(5)}}$}
\label{sec:obtaining-lagrangian-m5}
\setcounter{equation}{0}

Here it will be outlined how to obtain the Lagrangian for the dimension-5 part of the operator $\widehat{m}$, cf.~\eqref{eq:lagrangian-m5-complete}.
First, we will work in an observer frame with the mixed coefficients $m^{(5)}_{0i}\neq 0$ only. The Gr\"{o}bner basis is computed for the ordering
$L>\dots>p_0$ of the variables; it comprises 66 polynomials.
From these polynomials one has to find five suitable ones to solve the initial equations successively. A reasonable choice is such that the first
equation depends on $p_0$ only, the second on both $p_0$ and $p_1$, etc. Then the first equation can be solved with respect to $p_0$ and the second
with respect to $p_1$, etc. This leads to a chain of replacement rules that can be inserted into Euler's formula for the Lagrangian to obtain the
latter step by step:
\begin{subequations}
\begin{align}
(p_0)^{(0)}&=0\,, \\[2ex]
(p_0)^{(2,3)}&=\pm\frac{m_{\psi}\mathrm{sgn}(u^0)}{[(u^0)^2-\mathbf{u}^2]^{3/2}}\Big\{u^0\left[(u^0)^2-\mathbf{u}^2\right]-2m_{\psi}\mathbf{u}^2\sum_i m^{(5)}_{0i}u^i\Big\}\,, \\[2ex]
p_1&=\frac{p_0}{(u^0)^2}\left(2m_{\psi}\left\{m^{(5)}_{01}[(u^0)^2+(u^1)^2]+m^{(5)}_{02}u^1u^2+m^{(5)}_{03}u^1u^3\right\}-u^0u^1\right)\,, \\[2ex]
p_2&=\frac{p_0}{(u^0)^2}\left(2m_{\psi}\left\{m^{(5)}_{02}[(u^0)^2+(u^2)^2]+m^{(5)}_{03}u^2u^3\right\}-u^0u^2\right) \notag \\
&\phantom{{}={}}-2m_{\psi}m^{(5)}_{01}\frac{u^2}{u^0}p_1\,, \\[2ex]
p_3&=\frac{u^3}{u^2}\left(p_2-2m_{\psi}m^{(5)}_{02}p_0\right)+2m_{\psi}m^{(5)}_{03}p_0\,, \\[2ex]
L&=-(u^0p_0+u^1p_1+u^2p_2+u^3p_3)\,,
\end{align}
with the sign function:
\begin{equation}
\mathrm{sgn}(x)\equiv \left\{\begin{array}{rcl}
1 & \text{for} & x>0 \\
0 & \text{for} & x=0 \\
-1 & \text{for} & x<0\,. \\
\end{array}
\right.
\end{equation}
\end{subequations}
We obtain three solutions for $p_0$ where the first is trivial. After a final expansion in the Lorentz-violating component coefficients, the remaining
ones lead to the two possible Lagrangians stated in \eqref{eq:lagrangian-m5-mixed}.
To obtain \eqref{eq:lagrangian-m5-spatial} we choose an observer frame with the spatial component coefficients $m^{(5)}_{ij}\neq 0$. For that case
the Gr\"{o}bner basis is computed with respect to the ordering $p_0>\dots>L$. A single polynomial of this basis comprises $L$ only. Therefore, we
obtain the equation
\begin{align}
0&=\Big[1+2m_{\psi}\Big(\sum_i m^{(5)}_{ii}\Big)\Big]L^2-\Big(1+2m_{\psi}\sum_i m^{(5)}_{ii}\Big)m_{\psi}^2[(u^0)^2-\mathbf{u}^2] \notag \\
&\phantom{{}={}}+2m_{\psi}^3\sum_{i,j} m^{(5)}_{ij}u^iu^j\,.
\end{align}
Solving the latter with respect to $L$ and performing a successive expansion at first order in Lorentz violation leads to \eqref{eq:lagrangian-m5-spatial}.

This demonstrates both methods that can be used to obtain classical Lagrangians from Gr\"{o}bner bases. In comparison to the second method, the first
does not provide the solution directly. Nevertheless, the first is preferable for a number of reasons. Each of the equations obtained via the first
technique can be simplified and expanded separately whereas the second method may deliver a high-order polynomial that may possibly not be easy to
solve. Furthermore, the first method seems to deliver a Gr\"{o}bner basis faster with less polynomials. Hence, for all the remaining cases we will
proceed according to the first.

\section{Isotropic parts of the Lagrangians}
\label{sec:obtaining-isotropic-lagrangians}
\setcounter{equation}{0}

The current section shall outline the derivation of the isotropic Lagrangians in \secref{sec:isotropic-parts}. It is understood that equal indices are
not summed over.
Consider the operator $\widehat{a}^{(5)}_{\mathrm{eff}}$ whose isotropic component coefficients can be found in Eq.~(97) of~\cite{Kostelecky:2013rta}.
The first isotropic part is governed by the single coefficient $(a^{(5)}_{\mathrm{eff}})_{000}\equiv \ring{a}{}^{(5)}_0$ and all others
vanishing. For such a configuration we obtain
\begin{equation}
\widehat{a}^{(5)}_{\mathrm{eff,\ast}}|_{\ring{a}{}^{(5)}_0}=\ring{a}{}^{(5)}_0(u^0)^3\,.
\end{equation}
The second part is based on the coefficients $(a^{(5)}_{\mathrm{eff}})_{0jj}$ including their symmetric set of index permutations. There are nine
of such permutations, three for each $j$. Denoting each of these component coefficients with $\xi$ and setting the remaining ones to
zero leads to
\begin{equation}
\widehat{a}^{(5)}_{\mathrm{eff,\ast}}|_{\ring{a}{}^{(5)}_2}=3\xi u^0\mathbf{u}^2=\ring{a}{}^{(5)}_2u^0\mathbf{u}^2\,.
\end{equation}
The operator $\widehat{c}^{\,(6)}_{\mathrm{eff}}$ has three isotropic parts given in Eq.~(98) of~\cite{Kostelecky:2013rta}. The first is made up by a
single nonvanishing controlling coefficient $(c^{(6)}_{\mathrm{eff}})_{0000}\equiv \ring{c}{}^{(6)}_0$:
\begin{equation}
\widehat{c}^{\,(6)}_{\mathrm{eff},\ast}|_{\ring{c}{}^{(6)}_0}=\ring{c}{}^{(6)}_0(u^0)^4\,.
\end{equation}
The second is comprised by the coefficients $(c^{(6)}_{\mathrm{eff}})_{00jj}$ and related ones by symmetric index permutations. In total there are 18
of such permutations, six for each $j$. Denoting each of these coefficients with $\zeta$ where the others are supposed to vanish results in
\begin{equation}
\widehat{c}^{\,(6)}_{\mathrm{eff},\ast}|_{\ring{c}{}^{(6)}_2}=6\zeta (u^0)^2\mathbf{u}^2=\ring{c}{}^{(6)}_2(u^0)^2\mathbf{u}^2\,.
\end{equation}
There is a third isotropic part made up by the coefficients $(c^{(6)}_{\mathrm{eff}})_{jjkk}$ and related ones by symmetries. For each $j$
and $k$ with $j<k$ there are six permutations leading to 18 in total. Using the notation $(c^{(6)}_{\mathrm{eff}})_{jjkk}\equiv\psi$ for
$j\neq k$ and $(c^{(6)}_{\mathrm{eff}})_{jjjj}\equiv 3\psi$ with all others set to zero we have that
\begin{equation}
\widehat{c}^{\,(6)}_{\mathrm{eff},\ast}|_{\ring{c}{}^{(6)}_4}=3\psi \mathbf{u}^4=\ring{c}{}^{(6)}_4\mathbf{u}^4\,.
\end{equation}
The index structure of the component coefficients for $\widehat{e}^{\,(6)}$ and $\widehat{f}^{(6)}$ is the same as of $\widehat{a}^{\,(5)\mu}$, which
is why the isotropic parts can be derived analogously for these cases. These results are used to obtain the isotropic Lagrangians in
\secref{sec:isotropic-parts}.

\section{Cartan torsion}
\label{sec:cartan-matsumoto-torsion}
\setcounter{equation}{0}

For the generic Finsler structure of \eqref{eq:generic-finsler-structure}, the Cartan torsion is computed according to the definition given by
\eqref{eq:cartan-torsion-definition}. After arranging terms properly with respect to powers of $m_{\psi}$ and $y^2$ the result reads as follows:
\begin{subequations}
\label{eq:cartan-torsion-generic}
\begin{align}
\frac{2}{F}C_{ijk}&=\frac{m_{\psi}^{l_1}}{\sqrt{y^2}}\left(\frac{1}{y^2}\mathcal{C}_{ij}y_k-\mathcal{C}^{(1)}_{ijk}\right)+m_{\psi}^{2l_1}\mathcal{C}^{(2)}_{ijk}\,, \displaybreak[0]\\[2ex]
\mathcal{C}_{ij}&=\widehat{O}_{\ast}\bigg[f\bigg(\delta_{ij}-\frac{y_iy_j}{y^2}\bigg)+\bigg(y_i\frac{\partial f}{\partial y^j}+(i\leftrightarrow j)\bigg)+y^2\frac{\partial^2f}{\partial y^i\partial y^j}\bigg] \notag \\
&\phantom{{}={}\widehat{O}_{\ast}\bigg[}+\bigg[\frac{\partial\widehat{O}_{\ast}}{\partial y^j}\bigg(y^if+y^2\frac{\partial f}{\partial y^i}\bigg)+(i\leftrightarrow j)\bigg]+y^2f\frac{\partial^2\widehat{O}_{\ast}}{\partial y^i\partial y^j}\,, \displaybreak[0]\\[2ex]
\mathcal{C}^{(1)}_{ijk}&=\frac{\partial\widehat{O}_{\ast}}{\partial y^k}\bigg[f\bigg(\delta_{ij}-\frac{y_iy_j}{y^2}\bigg)+\bigg(y^i\frac{\partial f}{\partial y^j}+(i\leftrightarrow j)\bigg)+y^2\frac{\partial^2f}{\partial y^i\partial y^j}\bigg] \notag \\
&\phantom{{}={}}+\widehat{O}_{\ast}\bigg\{\frac{\partial f}{\partial y^k}\bigg[\delta_{ij}-\frac{y_iy_j}{y^2}\bigg]-\frac{f}{y^4}\bigg[y^2\big(\delta_{ik}y_j+(i\leftrightarrow j)\big)-2y_iy_jy_k\bigg] \notag \\
&\phantom{{}={}+\widehat{O}_{\ast}\bigg[}+\bigg[\delta_{ik}\frac{\partial f}{\partial y^j}+y_i\frac{\partial^2f}{\partial y^j\partial y^k}+(i\leftrightarrow j)\bigg]+2y_k\frac{\partial^2f}{\partial y^i\partial y^j}+y^2\frac{\partial^3f}{\partial y^i\partial y^j\partial y^k}\bigg\} \notag \\
&\phantom{{}={}}+\bigg[\frac{\partial^2\widehat{O}_{\ast}}{\partial y^j\partial y^k}\bigg(y^if+y^2\frac{\partial f}{\partial y^i}\bigg)+\frac{\partial\widehat{O}_{\ast}}{\partial y^j}\bigg(\delta_{ik}f+y_i\frac{\partial f}{\partial y^k}+2y_k\frac{\partial f}{\partial y^i}+y^2\frac{\partial^2f}{\partial y^i\partial y^k}\bigg) \notag \\
&\phantom{{}={}+\bigg[}+(i\leftrightarrow j)\bigg]+2y_kf\frac{\partial^2\widehat{O}_{\ast}}{\partial y^i\partial y^j}+y^2\frac{\partial f}{\partial y^k}\frac{\partial\widehat{O}_{\ast}}{\partial y^i\partial y^j}+y^2f\frac{\partial\widehat{O}_{\ast}}{\partial y^i\partial y^j\partial y^k}\,, \displaybreak[0]\\[2ex]
\mathcal{C}^{(2)}_{ijk}&=2\widehat{O}_{\ast}\frac{\partial\widehat{O}_{\ast}}{\partial y^k}\bigg(\frac{\partial f}{\partial y^i}\frac{\partial f}{\partial y^j}+f\frac{\partial^2f}{\partial y^i\partial y^j}\bigg) \notag \\
&\phantom{{}={}}+(\widehat{O}_{\ast})^2\bigg[\bigg(\frac{\partial f}{\partial y^i}\frac{\partial^2f}{\partial y^j\partial y^k}+(i\leftrightarrow j)\bigg)+\frac{\partial f}{\partial y^k}\frac{\partial^2f}{\partial y^i\partial y^j}+f\frac{\partial^3f}{\partial y^i\partial y^j\partial y^k}\bigg] \notag \\
&\phantom{{}={}}+2\bigg(\widehat{O}_{\ast}\frac{\partial f}{\partial y^k}+f\frac{\partial\widehat{O}_{\ast}}{\partial y^k}\bigg)\bigg[\frac{\partial f}{\partial y^i}\frac{\partial\widehat{O}_{\ast}}{\partial y^j}+(i\leftrightarrow j)\bigg]+2f\frac{\partial f}{\partial y^k}\bigg(\frac{\partial\widehat{O}_{\ast}}{\partial y^i}\frac{\partial\widehat{O}_{\ast}}{\partial y^j}+\widehat{O}_{\ast}\frac{\partial^2\widehat{O}_{\ast}}{\partial y^i\partial y^j}\bigg) \notag \\
&\phantom{{}={}}+2f\widehat{O}_{\ast}\bigg(\frac{\partial f}{\partial y^i}\frac{\partial\widehat{O}_{\ast}}{\partial y^j\partial y^k}+\frac{\partial^2f}{\partial y^i\partial y^k}\frac{\partial\widehat{O}_{\ast}}{\partial y^j}+(i\leftrightarrow j)\bigg) \notag \\
&\phantom{{}={}}+f^2\bigg(\frac{\partial\widehat{O}_{\ast}}{\partial y^i}\frac{\partial^2\widehat{O}_{\ast}}{\partial y^j\partial y^k}+(i\leftrightarrow j)+\frac{\partial\widehat{O}_{\ast}}{\partial y^k}\frac{\partial^2\widehat{O}_{\ast}}{\partial y^i\partial y^j}+\widehat{O}_{\ast}\frac{\partial^3\widehat{O}_{\ast}}{\partial y^i\partial y^j\partial y^k}\bigg)\,.
\end{align}
\end{subequations}
Finally, the generic result for the angular metric tensor that is needed for the Matsumoto torsion in \eqref{eq:matsumoto-torsion-definition} is
given by
\begin{align}
\label{eq:average-curvature-generic}
\frac{h_{ij}}{F}&=\frac{\delta_{ij}}{\sqrt{y^2}}-\frac{y_iy_j}{(y^2)^{3/2}}-m_{\psi}^{l_1}\bigg[\widehat{O}_{\ast}\frac{\partial^2f}{\partial y^i\partial y^j}+\bigg(\frac{\partial f}{\partial y^i}\frac{\partial \widehat{O}_{\ast}}{\partial y^j}+(i\leftrightarrow j)\bigg)+f\frac{\partial^2\widehat{O}_{\ast}}{\partial y^i\partial y^j}\bigg]\,.
\end{align}

\end{appendix}

\newpage


\end{document}